\documentclass[prb,twocolumn,aps,amsmath,amssymb]{revtex4}
\usepackage[T1]{fontenc}
\usepackage[utf8]{inputenc}
\usepackage{graphicx}
\usepackage{amsmath}
\usepackage{amssymb}
\usepackage{dcolumn}
\usepackage{float}
\usepackage{bm}
\usepackage{amsfonts,times}
 \usepackage[breaklinks=true,colorlinks,citecolor=blue,linkcolor=blue,urlcolor=blue]{hyperref}

\newcommand{\eins}{\mbox{$1 \hspace{-1.0mm} {\bf l}$}}

\DeclareMathAlphabet{\bi}{OML}{cmm}{b}{it}

\def\be{\begin{equation}}
\def\ee{\end{equation}}
\def\bearr{\begin{eqnarray}}
\def\eearr{\end{eqnarray}}

\def\bs{\boldsymbol}

\begin{document}
\title{Floquet engineering of low-energy dispersions and dynamical localization in a  periodically kicked {three-band system}}
\bigskip

\author{Lakpa Tamang$\color{blue}{^1}$}
\author{Tanay Nag$\color{blue}{^{2}}$}
\email{tnag@physik.rwth-aachen.de}
\author{Tutul Biswas$\color{blue}{^1}$}
\email{tbiswas@nbu.ac.in}
\normalsize
\affiliation
{$\color{blue}{^1}$Department of Physics, University of North Bengal, Raja Rammohunpur-734013, India\\
$\color{blue}{^{2}}$ 
Institut f\"ur Theorie der Statistischen Physik, RWTH Aachen University, 
52056 Aachen, Germany}
\date{\today}

\begin{abstract}
Much having learned about Floquet dynamics of pseudospin-$1/2$ system namely, graphene, we  
here address the
stroboscopic properties of a periodically kicked {three-band fermionic system such as 
$\alpha$-T$_3$ lattice. This particular model provides an interpolation between graphene and dice lattice via the continuous tuning of the parameter $\alpha$ from 0 to 1.} In the case of dice lattice ($\alpha=1$),
we reveal that one can, in principle, engineer various types of 
low energy dispersions around some specific points in the Brillouin zone by tuning the kicking
parameter in the Hamiltonian
along a particular direction. Our analytical analysis shows that one can experience different 
quasienergy dispersions for example,
Dirac type, semi-Dirac type, gapless line, absolute flat quasienergy bands,  depending
on the specific values of the kicking parameter. Moreover, we numerically study the dynamics of a  
wave packet in dice lattice. The quasienergy dispersion allows us to understand the instantaneous structure of  wave packet at stroboscopic times. We find a situation where absolute flat quasienergy bands lead to a complete dynamical localization of the wave packet. {Aditionally, we calculate the quasienergy spectrum numerically for 
$\alpha$-T$_3$ lattice. A periodic kick in a perpendicular (planar) direction breaks (preserves) the particle-hole symmetry for $0<\alpha<1$.  
Furthermore, it is also revealed that the dynamical localization of wave packet does not occur at any intermediate $\alpha \ne 0,\,1$.}
\end{abstract}

\maketitle

\section{Introduction}
Graphene[\onlinecite{Grph_dis}], a strictly two dimensional sheet of carbon atoms arranged on a honeycomb lattice, 
brings a new revolution in condensed matter physics
due to its fascinating physical properties [\onlinecite{grphn_rev1, grphn_rev2, grphn_rev3}]  and its potential application
in the field of nanotechnology[\onlinecite{grphn_tran}].
The itinerant electrons in graphene
behave like quasiparticles of which the low energy massless excitations are described by the pseudospin-$1/2$ Dirac-Weyl equation.
Thus, graphene provides a platform to study the fingerprints of the Dirac physics in realistic systems.

There exists an analogous model with $T_3$-symmetry, known as the dice lattice model[\onlinecite{dice1,dice2}],
which also exhibits low energy massless 
excitations. The geometry of the dice lattice consists of an additional site being located at the center of each hexagon 
of the honeycomb lattice and that additional site is connected to one of the two  inequivalent sites 
of the honeycomb lattice. A dice lattice can be found in a trilayer structure[\onlinecite{dice_grow}] of cubic lattices, namely,
SrTiO$_3$/SrIrO$_3$/SrTiO$_3$ grown in $(111)$-direction. It is also proposed that one can realize a dice lattice model in an 
optical lattice[\onlinecite{dice_opt}] by confining cold atoms using three pairs of counter propagating laser beams.
A slightly modified lattice, named as,
$\alpha$-T$_3$ lattice[\onlinecite{alph_T3}] is also an important topic of current research as it plays 
a role of interpolating between graphene and dice
lattice through the variation of the parameter $\alpha\in[0,1]$ and this variation is associated with distinct non-trivial
Berry phase. In recent years, a number of studies have been performed in dice and $\alpha$-T$_3$ lattice from the point of view of  {local topology induced localization}[\onlinecite{dice1,dice2}], 
magnetic frustration[\onlinecite{frust1,frust2}], spin-orbit interaction induced phenomena[\onlinecite{dice_SOI1}], 
Klein tunneling[\onlinecite{dice_Klein, alp_T3_Klein}], plasmon[\onlinecite{plasm, plasm2}], 
magneto-optical conductivity[\onlinecite{dice_Berry, dice_MagOP1,dice_MagOP2}], magnetotransport[\onlinecite{alp_T3Mag}], 
spatial modulation effect[\onlinecite{alp_T3Mod}],
zitterbewegung[\onlinecite{alp_T3Zb}], non-linear optical response[\onlinecite{alp_nonOp}],
RKKY interaction[\onlinecite{RKKY1, RKKY2}], minimal conductivity[\onlinecite{alp_min}], 
geometric quench[\onlinecite{gm_qnch}],
topological phases in a Haldane-dice 
lattice model[\onlinecite{Haldane_Dice}] etc.

Although both graphene and dice lattice share same zero field spectrum, nevertheless, these two systems are fundamentally different {as the later system hosts a zero energy flat band}.
In graphene the pseudospin of a quasiparticle is $S=1/2$ whereas the dice lattice hosts quasiparticles with pseudospin $S=1$. In 
an external magnetic field, graphene behaves like a diamagnet[\onlinecite{Grph_dia}] while
a dice lattice model exhibits paramagnetic[\onlinecite{alph_T3}] response. The 
Hall quantization[\onlinecite{dice_Berry, alp_T3Mag}] rules are also different in these two systems. 
The quasiparticles in graphene acquire a non-trivial Berry phase 
of $\pi$ while traversing a closed loop around a high symmetry point in momentum space. In contrary, the quasiparticles in a 
dice lattice model do not pick any non-trivial Berry phase during such movement.

Recent years have witnessed a tremendous quest to understand the microscopic details of quantum systems driven by 
external time periodic fields. The Floquet theory[\onlinecite{Floquet1, Floquet2}] provides an extremely 
useful theoretical framework to deal with the time 
periodic Hamiltonians corresponding to the driven systems. 
The studies[\onlinecite{Gr_Fl1, Gr_Fl2, Gr_Fl3, Gr_Fl4}] on graphene irradiated 
by the circularly polarized time periodic fields have become so impactful that an exciting research 
field of ``Floquet topological insulators'' has been emerged subsequently[\onlinecite{Fl_topo1, Fl_topo2,Fl_topo3}].
Over a period of last ten years, Floquet irradiated quantum systems have been explored extensively in the context of
Floquet generation of strongly correlated phases[\onlinecite{Fl_Cor1, Fl_Cor2, Fl_Cor3, Fl_Cor4, Fl_Cor5}],
symmetry protected topological phases[\onlinecite{SmPTp1,SmPTp2,SmPTp3,SmPTp4,SmPTp5}] 
in many-body quantum systems,
topological classification[\onlinecite{TCls1, TCls2, TCls3, TCls4, TCls5, TCls6, TCls7,TCls8, TCls9}], 
symmetry breaking[\onlinecite{Sym1, Sym2, Sym3, Sym4, Sym5, Sym6, Sym8}], Floquet-Majorana
modes[\onlinecite{FMaj1,FMaj2,FMaj3, FMaj4, FMaj5,FMaj6}], topologically protected edge 
states[\onlinecite{Edg2, Edg3, Edg4, Edg5}], 
Floquet topological phase transition[\onlinecite{FTP1,FTP2}], Floquet flat band [\onlinecite{FFB}] etc. It is worthy to mention that many of these 
phenomena have been realized experimentally[\onlinecite{Exp1, Exp2, Exp3, Exp4, Exp5}] in recent past.

To study the stroboscopic
properties of a driven quantum system, driving protocols like the $\delta$-function kicks, periodic in time, may be adopted. The 
impact of such driving protocol has far reaching consequences. This type of driving has been used to study a number of spectacular phenomena including non-equilibrium phase transition in a Dicke Model[\onlinecite{Dicke}],
localization effect in a chain 
of hard core bosons[\onlinecite{loc_HCB}], semimetallic phases in Harper models[\onlinecite{Harper}], 
edge modes in quantum Hall 
systems[\onlinecite{edge_QHS}], low energy band engineering in graphene[\onlinecite{Graphn_band}],
Majorana edge mode in one dimensional systems[\onlinecite{Majrn1d}] as well as in Kitaev model[\onlinecite{Thaku_Kit}] on
a honeycomb lattice,
topological properties of Chern insulator[\onlinecite{Chern_topo}], topological phase transition in
Haldane-Chern insulator[\onlinecite{HChrn_T}], 
generation of higher order topological insulator from a lower order topological insulating phase [\onlinecite{SmPTp6}] and many more.
Another interesting effect of periodic $\delta$-function 
driving on the quantum systems is to achieve dynamical localization of the quasiparticles. A number of systems like classical 
and quantum rotors[\onlinecite{rot1,rot2,rot3, rot4}], 
two level system[\onlinecite{TLS, TLS2}], the Kapitza pendulum[\onlinecite{Kap1, Kap2}], bosons in optical lattice[\onlinecite{Opt_Boson}],
linear chain of hard core bosons[\onlinecite{loc_HCB}], graphene[\onlinecite{Graphn_band}] etc
exhibit dynamical localization phenomenon under periodic driving.

Given this background, one can comment that the study of Floquet dynamics in the 
{three-band} system has not been addressed in great detail as it is investigated for {two-band} systems. The transport properties
of a {three-band} system can become significantly different from that of a {two-band} system, specially while the former 
system hosts a flat band. This also motivates us to investigate the {three-band 
$\alpha$-T$_3$ lattice model} that has the
potential to exhibit intriguing band dispersions eventually leading to unusual non-equilibrium transport properties. 
In particular, we study the
stroboscopic non-equilibrium properties of a periodically kicked  { $\alpha$-T$_3$ lattice. For $0<\alpha<1$, it is not possible to handle this problem analytically. Although analytical results are possible to obtain in the case of a dice lattice corresponding to $\alpha=1$. For the dice lattice,}  we check that the characteristics of the 
zero-energy flat band  remains unaltered in presence of the $\delta$-function kicks. It is revealed that various types
of low energy dispersions including Dirac type, semi-Dirac type, gapless line, absolute flat quasienergy bands can be 
engineered around some special points in the Brillouin zone. This wide variety of dispersions entirely depends on the
tuning of the kicking parameter and direction of periodic kicking. In addition, we study wave packet dynamics in this system. The periodic $\delta$-driving along the transverse direction causes the dynamical localization of the electronic wave packet corresponding 
to a certain strength of kicking parameter. The origin of this dynamical localization lies in the fact that all
three quasienergy bands become absolutely flat 
at a certain value of the kicking strength.
For kicking in other directions, we obtain diffusive character of the wave packet as we do not encounter a situation where all the three quasienergy bands become flat.
{ The quasienergy spectrum 
of a periodically kicked $\alpha$-T$_3$ lattice
is obtained numerically for various kicking direction. It is obtained that periodic kicking in a perpendicular direction lifts the particle-hole symmetry while an in-plane kicking
respects that symmetry. It is also understood that the dynamical localization phenomenon of wave packet is absent for all values of $\alpha$ except for
$\alpha=0$ (graphene) [\onlinecite{Graphn_band}] and $\alpha=1$ (dice lattice).}

This article is presented in the following way. In section II, we provide a brief description of the geometry and energy spectrum 
of the { $\alpha$-T$_3$} lattice. {A brief description of Floquet theory is presented in section III.} Detailed characteristics of the quasienergy spectrum of a periodically kicked dice lattice corresponding 
to different kicking schemes are described in section IV. Different aspects of the wave packet dynamics {in dice lattice} are investigated in section V. {Numerical results for a driven 
$\alpha$-T$_3$ lattice model have been provided in section VI}.
Finally, we conclude our findings in section VII.

\section{A description of the $\alpha$-T$_3$ lattice}

{An $\alpha$-T$_3$ lattice} has a bipartite honeycomb like structure with an additional site at the center of 
each hexagon.
As shown in Fig. 1 (Upper Panel), $A$ and $B$ atoms form the honeycomb structure with nearest neighbor 
hopping amplitude $\gamma$. The $C$ atom sitting at the center of each hexagon is connected to the $B$ atoms 
with hopping parameter {$\alpha\gamma$, where the parameter $\alpha$ can take any value in between $0$ and $1$. Thus $\alpha=0\,(1)$ corresponds to graphene\,(dice lattice).}  Both $A$ and $C$ atoms are connected to the three $B$ atoms, hence 
both have coordination number $3$, named as rim sites. The $B$ atom has the coordination number $6$, known as 
hub site. Each nearest neighbor pair consists of one hub atom and one rim atom. Each unit cell contains 
three lattice sites. The lattice \textcolor{black}{translation} vectors are ${\bm a}_1=(\sqrt{3}a/2,3a/2)$
and ${\bm a}_2=(-\sqrt{3}a/2,3a/2)$, where $a$ is the lattice constant. Therefore, the entire lattice is spanned by ${\bm n}=n_1{\bm a}_1+n_2{\bm a}_2$, 
$n_1,n_2\in Z$. The coordinates of $A$ sites surrounded by the $B$ site are 
${\bs \delta}_A^1=(0,-a)$, ${\bs \delta}_A^2={\bs \delta}_A^1+{\bm a}_1$, and ${\bs \delta}_A^3={\bs \delta}_A^1+{\bm a}_2$.
The coordinates of $C$ sites surrounded by the $B$ site are are ${\bs \delta}_C^i=-{\bs \delta}_A^i$, $i=1,2,3$.
The reciprocal lattice vectors are ${\bm b}_1=b(1/\sqrt{3},1/3)$ and ${\bm b}_2=b(-1/\sqrt{3},1/3)$, where $b=2\pi/a$.

\begin{figure}[h!]
\begin{minipage}[b]{\linewidth}
\centering
 \includegraphics[width=5.5cm, height=4.5cm]{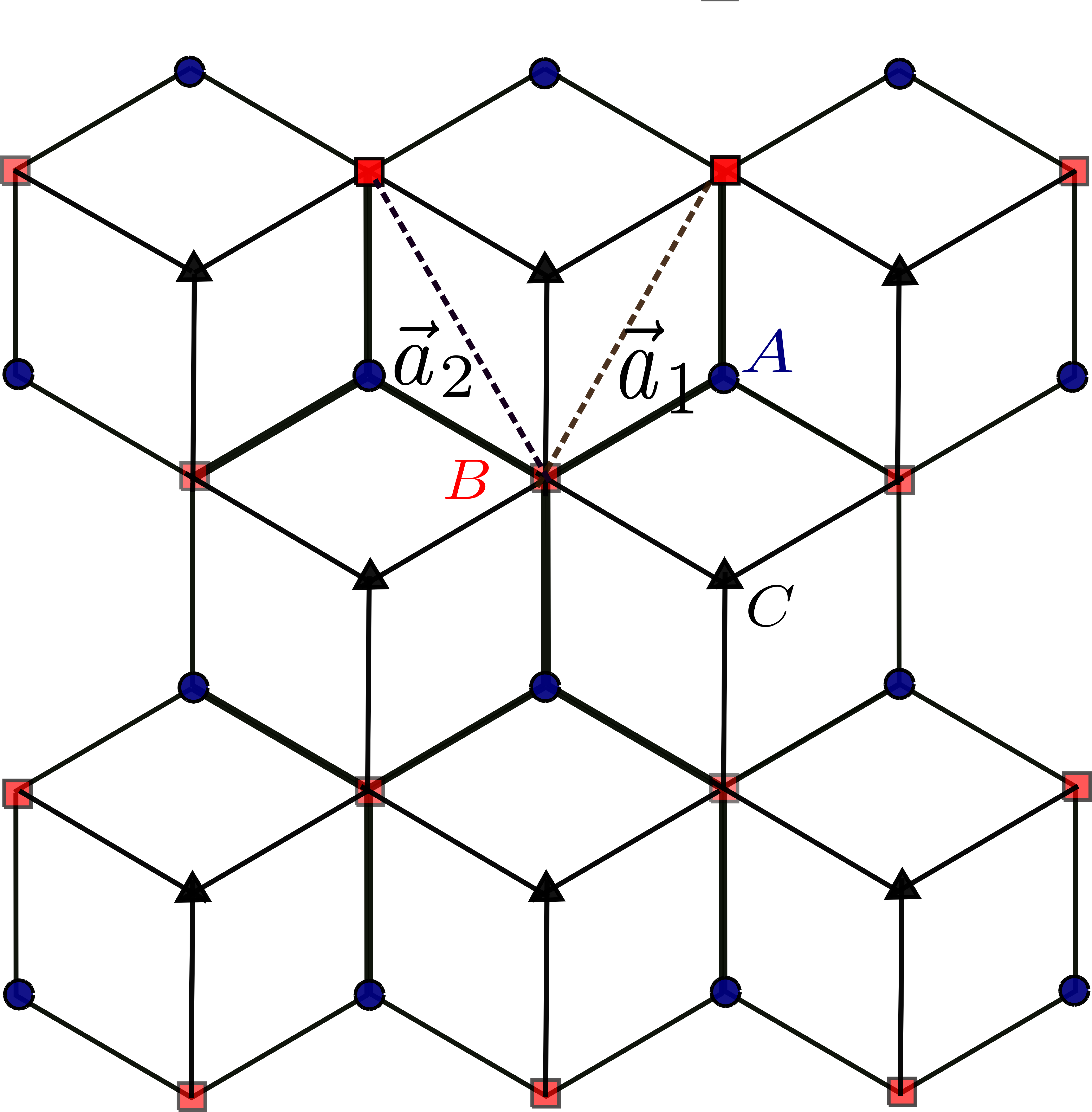}
\end{minipage}
\vspace{1em}

\begin{minipage}[b]{\linewidth}
\centering
\includegraphics[width=6.5cm, height=5cm]{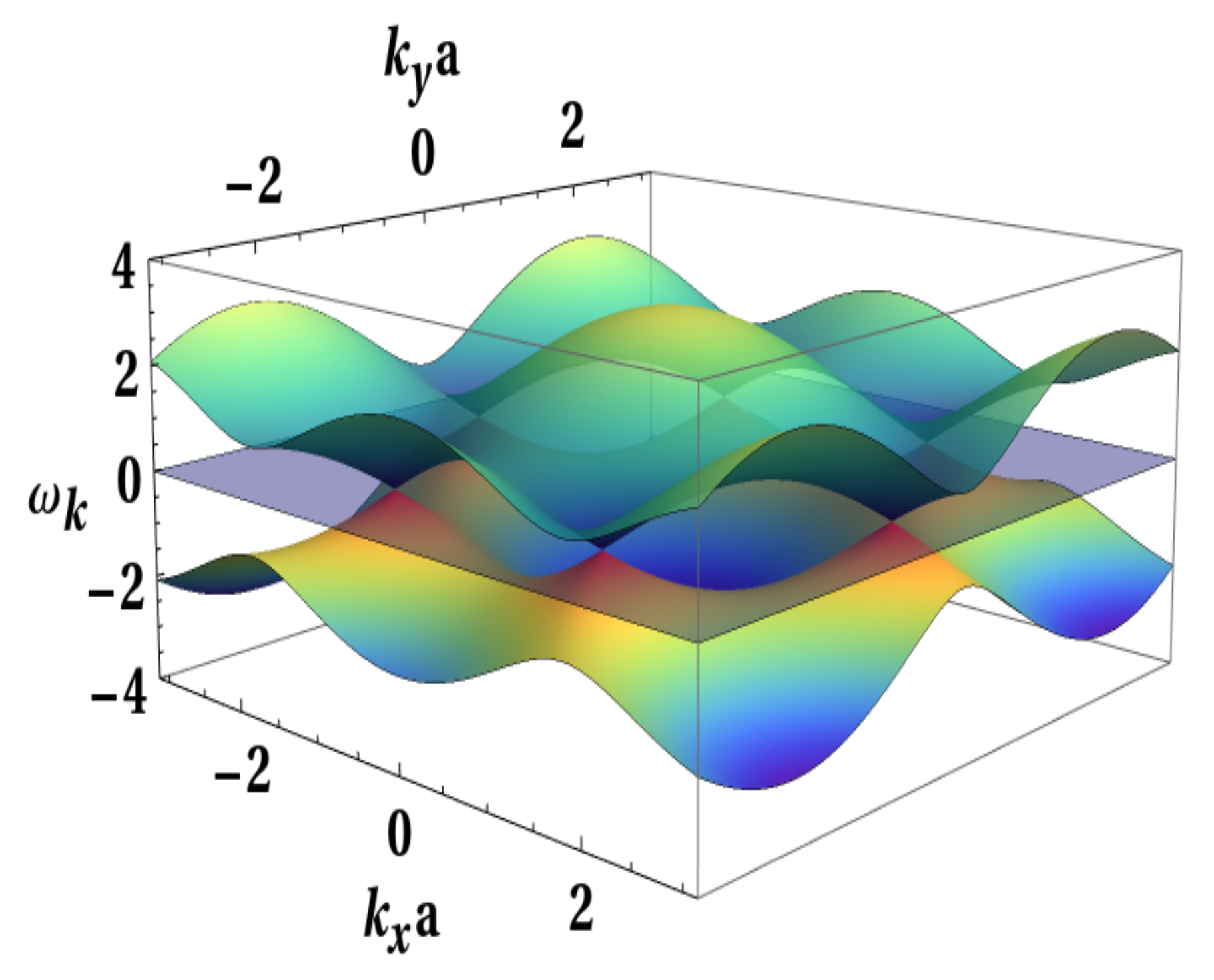}
\end{minipage}

\caption{In the upper panel the geometrical structure of \textcolor{black}{$\alpha$-T$_3$} lattice is shown. The energy spectrum is given in the lower 
panel obtained from Eq.(\ref{Nokick}).}
\end{figure}

The nearest neighbor tight-binding Hamiltonian can be written as 
\begin{eqnarray}\label{Ham1}
 H&=&-\gamma\sum_{{\bm n},{\bs \delta}_A}\Big(\mathcal{B}_{\bm n}^\dagger
 \mathcal{A}_{{\bm n}+{\bs \delta}_A}+\mathcal{A}_{{\bm n}+{\bs \delta}_A}^\dagger 
 \mathcal{B}_{\bm n}\Big)\nonumber\\
&-&\alpha\gamma\sum_{{\bm n},{\bs \delta}_C}\Big(\mathcal{B}_{\bm n}^\dagger
\mathcal{C}_{{\bm n}+{\bs \delta}_C}+\mathcal{C}_{{\bm n}+{\bs \delta}_C}^\dagger \mathcal{B}_{\bm n}\Big),
\end{eqnarray}
where $\mathcal{A}(\mathcal{A}^\dagger)$,
$\mathcal{B}(\mathcal{B}^\dagger)$, and $\mathcal{C}(\mathcal{C}^\dagger)$ are
the annihilation(creation) operators for the sites $A$, $B$, and $C$,
respectively. Using the appropriate Fourier transformations the Hamiltonian in momentum space can be obtained as
\textcolor{black}{
\begin{eqnarray}\label{Ham_alpha}
H_{\bm k}^\alpha=
\begin{pmatrix}
0      &  f_{\bm k}\cos\phi &   0 \\
f_{\bm k}^\ast\cos\phi  &  0  &  f_{\bm k}\sin\phi \\
0     &  f_{\bm k}^\ast\sin\phi  &  0
\end{pmatrix},
\end{eqnarray}
where $f_{\bm k}=-\gamma \big(1+e^{-i{\bm k}\cdot {\bf a}_1}+e^{-i{\bm k}\cdot {\bf a}_2}\big)$ and  $\alpha=\tan\phi$. 
Here, the Hamiltonian $H_{\bm k}^\alpha$ is rescaled by $\cos\phi$.}
Diagonalizing $H_{\bm k}^\alpha$, one can obtain the energy spectrum as
$\varepsilon_{\bm k}^\pm=\pm \omega_{\bm k}$ and $\varepsilon_{\bm k}^0=0$, where $\omega_{\bm k}$ is given by
\begin{eqnarray}\label{Nokick}
\omega_{\bm k}&=&\gamma\Bigg[3+2\cos(\sqrt{3}ak_x)\nonumber\\
&+&4\cos\Big(\frac{\sqrt{3}}{2}ak_x\Big)
\cos\Big(\frac{3}{2}ak_y\Big)\Bigg]^{\frac{1}{2}}
\end{eqnarray}
and the eigenstates $|\psi_{\pm}^{\alpha}(\bm k)\rangle $ and  $|\psi_{0}^{\alpha}(\bm k)\rangle $. 
Here, we have chosen $\hbar=1$
and henceforth this choice will be maintained throughout this article. \textcolor{black}{The dispersive energy bands i.e. $\varepsilon_{\bm k}^\pm=\pm \omega_{\bm k}$ are identical to the energy bands of graphene. Aditionally, the static $\alpha$-T$_3$ lattice possesses a zero energy flat band. We note that the parameter $\alpha$ does not appear in energy dispersions, however the corresponding energy eigenstates will be $\alpha$ dependent.}
\textcolor{black}{We would like to note that $\alpha$-T$_3$ lattice respects particle-hole symmetry: $\varepsilon_{\bm k}^+=-\varepsilon_{\bm k}^-$ for any value of $\alpha$ within $0<\alpha\le 1$ while the associated wave-functions $|\psi_{+}^{\alpha}(\bm k)\rangle $ and $|\psi_{-}^{\alpha}(- \bm k)\rangle $ are related to each other by particle-hole symmetry ${\mathcal P}$: ${\mathcal P} |\psi_{\pm}^{\alpha}(\bm k)\rangle = |\psi_{\mp}^{\alpha}(-\bm k)\rangle$. For compactness, we provide the specific form of  the antiunitary particle-hole symmetry that is generated by ${\mathcal P} H_{\bm k}^{\alpha} {\mathcal P}^{-1}= -H_{-\bm k}^{\alpha}$ with ${\mathcal P} ={\mathcal D}_z {\mathcal K} $ and 
\begin{equation}\label{ph_operator}
 {\mathcal D}_z=\begin{pmatrix}
1 & 0 & 0\\
0 & -1 & 0\\
0 & 0 & 1
\end{pmatrix},
\end{equation}
and ${\mathcal K}$ denotes the complex conjugation operation. 
 Importantly, the particle-hole symmetry is also preserved for graphene where $\alpha=0$: ${\mathcal P} H_{\bm k}^{\alpha=0} {\mathcal P}^{-1}= -H_{-\bm k}^{\alpha=0}$ with ${\mathcal P}= \sigma_z {\mathcal K}$.
We will discuss the consequence of particle-hole symmetry in the context of driven system below.}

\section{Floquet theory}
Let us now consider the situation when the system is subjected to time periodic $\delta$-kicks of period $T$ in $x$, $y$, and 
$z$-directions with strengths $\lambda_x$, $\lambda_y$, and $\lambda_z$, respectively. This kicking process is described by 
the following Hamiltonian
\begin{eqnarray}\label{Kick_Ham}
V(t)=\sum_{\nu=x,y,z} \lambda_\nu S_\nu^\alpha \sum_{m=-\infty}^\infty \delta(t-mT),
\end{eqnarray}
\textcolor{black}{where the components of the pseudospin operator
$S_\nu^\alpha$($\nu=x,y,z$) are given by 
\begin{eqnarray}\label{Spin_alpha}
 &&S_x^\alpha=\begin{pmatrix}
0 & \cos\phi & 0\\
\cos\phi & 0 & \sin\phi\\
0 & \sin\phi & 0
\end{pmatrix},\nonumber\\
&&S_y^\alpha=\begin{pmatrix}
0 & -i\cos\phi & 0\\
i\cos\phi & 0 & -i\sin\phi\\
0 & i\sin\phi & 0
\end{pmatrix},\nonumber\\
 &&S_z^\alpha=2\begin{pmatrix}
\cos^2\phi & 0 & 0\\
0 & -\cos(2\phi) & 0\\
0 & 0 & -\sin^2\phi
\end{pmatrix}.
\end{eqnarray}}
We assume that all the unit cells are 
subjected to equal driving. We are mainly interested in the stroboscopic behavior of the system 
measured at the end of each driving period. The Floquet theory would thus be the best tool to handle \textcolor{black}{such} situation.
\textcolor{black}{The Floquet operator $U_F^\alpha(\bm k, T)$ describing the time evolution of the quantum states through one period $T$ is given by $U_F^\alpha(\bm k, T)=e^{-i{\bs \lambda}\cdot{\bf S}^\alpha} e^{-iH_{\bm k}^\alpha T}$. 
According to the Floquet theorem, the eigenvalues
and eigenvectors
of $U_F^\alpha(\bm k, T)$ are $e^{i \mu_n^\alpha(\bm k) T }$ and $|\phi_{n}^\alpha(\bm k)\rangle$, 
where $\mu_n^\alpha(\bm k)= \{ \Delta_{\bm k}^{\alpha \pm}, \Delta_{\bm k}^{\alpha 0} \}$ and  $|\phi_{n}^\alpha(\bm k)\rangle=\{ |\phi_{\pm}^\alpha(\bm k)\rangle, |\phi_{0}^\alpha(\bm k)\rangle \}$
are known as the quasienergies and quasi-states for $n=1,~2,~3$.
For $0<\alpha<1$, it is difficult to obtain the analytical expressions of the quasienergies corresponding to the periodic kicking in different directions. Therefore, we find the quasienergy spectrum numerically for an intermediate $\alpha$ and the corresponding results will be presented in section VI. However, the quasienergies of a kicked dice lattice ($\alpha=1$) can be obtained analytically to unveil the possibility of low energy band engineering. Hence, we first present the analytical results in next section.}  
\textcolor{black}{Before we proceed to section IV, we discuss the fate of particle-hole symmetry for the driven system.
The important role, played by particle-hole symmetry, can be directly manifested in the quasi-energy spectrum. 
For the driven case, the particle-hole symmetry is designated by ${\mathcal P} U_F^{\alpha}({\bm k},T) {\mathcal P}^{-1}= U_F^{\alpha}(-{\bm k},T)$ referring  to the fact that $\Delta_{\bm k}^{\alpha\pm} =- \Delta_{\bm k}^{\alpha \mp} $ and  ${\mathcal P} |\phi_{\pm}^\alpha(\bm k)\rangle = |\phi_{\mp}^\alpha(-\bm k)\rangle$ [\onlinecite{roy17}].}

\section{Quasienergy dispersion of a kicked Dice Lattice}
In this section, we consider a dice lattice subjected to the periodic driving according to the protocol described in Eq.(\ref{Kick_Ham}).
The explicit analytical expression of  quasienergy spectrum i.e. $\Delta_{\bm k}$ can be obtained by finding the energy eigenvalues of the Floquet operator 
$U_F(T)=e^{-i{\bs \lambda}\cdot{\bf S}} e^{-iH_{\bm k}T}$. \textcolor{black}{Here, the Hamiltonian $H_{\bm k}$ and the pseudospin operator ${\bm S}$ can be found from Eq.(\ref{Ham_alpha}) and Eq.(\ref{Spin_alpha}), respectively by setting $\alpha=1$ or equivalently $\phi=\pi/4$. It is too cumbersome to compute the quasienergies by considering all the kicking simultaneously.
Therefore, we will consider cases of individual kicking separately.}

\subsection{X-Kicking}
With the particular choice of kicking parameter, namely, $\lambda_x\neq0$, $\lambda_y=0$, and $\lambda_z=0$, the Floquet operator
becomes $U_F^x(T)=e^{-i\lambda_x S_x} e^{-iH_{\bm k}T}$. The 
eigenvalues of $U_F^x(T)$ can be obtained in a straightforward manner in order to calculate
the corresponding quasienergy $\Delta_{\bm k}^x$ (See Appendix for a detail derivation).
We obtain the quasienergies as
\begin{eqnarray}\label{XKick}
\Delta_{\bm k}^{x0}=0,~~~
\Delta_{\bm k}^{x\pm}=\pm \Delta_{\bm k}^x=\pm \frac{1}{T} \cos^{-1}(\kappa^x),
\end{eqnarray}
where
\begin{eqnarray}
\kappa^x&=&\frac{1}{2}\Big(\cos\lambda_x-1\Big)\sin^2\theta_{\bm k}-\sin\lambda_x
 \cos\theta_{\bm k}\sin(\omega_{\bm k}T)\nonumber\\
&+& \frac{1}{2}\Big[\sin^2\theta_{\bm k}+\cos\lambda_x\big(1+\cos^2\theta_{\bm k}\big)\Big]\cos(\omega_{\bm k}T).
\end{eqnarray}

It is noteworthy that the quasienergy spectrum contains one flat band and two dispersive bands. In other words, we may say 
that the fate of the zero energy flat band of the unkicked system remains unaltered. Similar feature is also obtained for a dice
lattice illuminated by circularly polarized radiation in the terahertz regime[\onlinecite{Sym8}]. 
Nevertheless, the external kicking modifies the \textcolor{black}{characteristics} of the dispersive bands significantly. 

To analyze the characteristics of the quasienergy spectrum we begin by writing $f_{\bm k}$ as $f_{\bm k}=-\gamma\big(\xi_{\bm k}-i\delta_{\bm k}\big)$ which enable us to express
$\cos\theta_{\bm k}$ and $\sin\theta_{\bm k}$ as:
$\cos\theta_{\bm k}=-\gamma \xi_{\bm k}/\omega_{\bm k}$ and $\sin\theta_{\bm k}=\gamma \delta_{\bm k}/\omega_{\bm k}$ with
$\omega_{\bm k}=\sqrt{\xi_{\bm k}^2+\delta_{\bm k}^2}$. Here, 
\begin{eqnarray}
&&\xi_{\bm k}=1+2\cos\Bigg(\frac{\sqrt{3}k_xa}{2}\Bigg)\cos\Bigg(\frac{3k_ya}{2}\Bigg)\nonumber\\
&&\delta_{\bm k}=2\cos\Bigg(\frac{\sqrt{3}k_xa}{2}\Bigg)\sin\Bigg(\frac{3k_ya}{2}\Bigg).
\end{eqnarray}
At gapless points $\kappa^x$ should be unity so that $\Delta_{\bm k}^{x\pm}=0$. This will be possible only when the 
following conditions are satisfied simultaneously: $\sin\theta_{\bm k}=0$, $\cos\theta_{\bm k}=-1$, and 
$\lambda_x=\omega_{\bm k}T$. 

Therefore, for the gapless points $(k_x^0, k_y^0)$ we have the following conditions
\begin{eqnarray}
&&1+2\cos\Bigg(\frac{\sqrt{3}k_x^0a}{2}\Bigg)\cos\Bigg(\frac{3k_y^0a}{2}\Bigg)=\frac{\lambda_x}{\gamma T}\nonumber\\
&&\cos\Bigg(\frac{\sqrt{3}k_x^0a}{2}\Bigg)\sin\Bigg(\frac{3k_y^0a}{2}\Bigg)=0
\end{eqnarray}
which further give
\begin{eqnarray}
&&\cos\Bigg(\frac{\sqrt{3}k_x^0a}{2}\Bigg)=\frac{1}{2}\Bigg(\frac{\lambda_x}{\gamma T}-1\Bigg)\nonumber\\
&&\sin\Bigg(\frac{3k_y^0a}{2}\Bigg)=0.
\end{eqnarray}

It is worthy to note that the value of $\lambda_x$ must lie in between $-\gamma T$ and $3\gamma T$ in order to obtain gapless points. To see how the quasienergy spectrum behaves around the gapless points $(k_x^0, k_y^0)$, we approximate 
$\cos\theta_{\bm k}$ and $\sin\theta_{\bm k}$ as: $\cos\theta_{\bm k}\approx-[1-\delta_{\bm k}^2/(2\xi_{\bm k}^2)]$ and 
$\sin\theta_{\bm k}\approx\delta_{\bm k}/\xi_{\bm k}$. With this approximation, we obtain following equation for 
the quasienergy spectrum
\begin{eqnarray}
\cos\big(\Delta_{\bm k}^xT\big)&=&\Big(1-\frac{\delta_{\bm k}^2}{2\xi_{\bm k}^2}\Big)\cos(\omega_{\bm k}T-\lambda_x)\nonumber\\
&+&\frac{\delta_{\bm k}^2}{2\xi_{\bm k}^2}\big[\cos(\omega_{\bm k}T)+(\cos\lambda_x-1)\big].
\end{eqnarray}

Expanding the sine and cosine functions about the gapless point, we get
\begin{eqnarray}
&&(\Delta_{\bm k}^xT)^2=(\omega_{\bm k}T-\lambda_x)^2+\frac{2\delta_{\bm k}^2}{\xi_{\bm k}^2}(1-\cos\lambda_x).
\end{eqnarray}

In the vicinity of the gapless points, we set $\xi_{\bm k}=\lambda_x/(\gamma T)$ to obtain the 
velocity components of the quasiparticle
\begin{eqnarray}
&&v_x=-\frac{\sqrt{3}a}{2T}\sqrt{3\gamma^2 T^2+2\gamma T\lambda_x-\lambda_x^2}\nonumber\\
&&v_y=\frac{3a}{T}\frac{(\lambda_x-\gamma T)}{\lambda_x}\sin\Big(\frac{\lambda_x}{2}\Big). 
\end{eqnarray}

\begin{figure}[h!]
 \begin{minipage}[b]{\linewidth}
\centering
 \includegraphics[width=6.5cm, height=5.5cm]{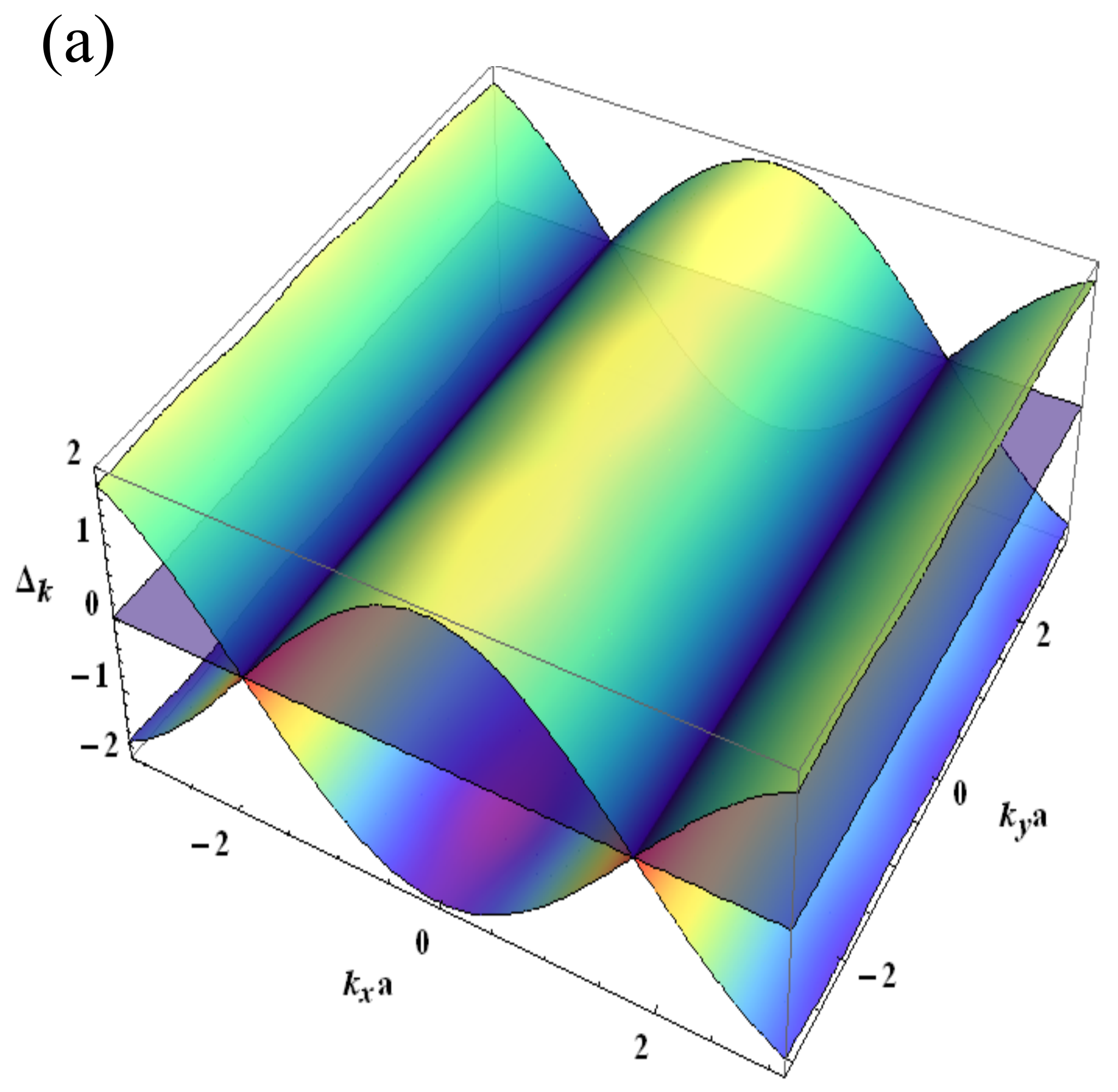}
 \end{minipage}
\vspace{0.1em}

\begin{minipage}[b]{\linewidth}
 \centering
 \includegraphics[width=6.5cm, height=5.5cm]{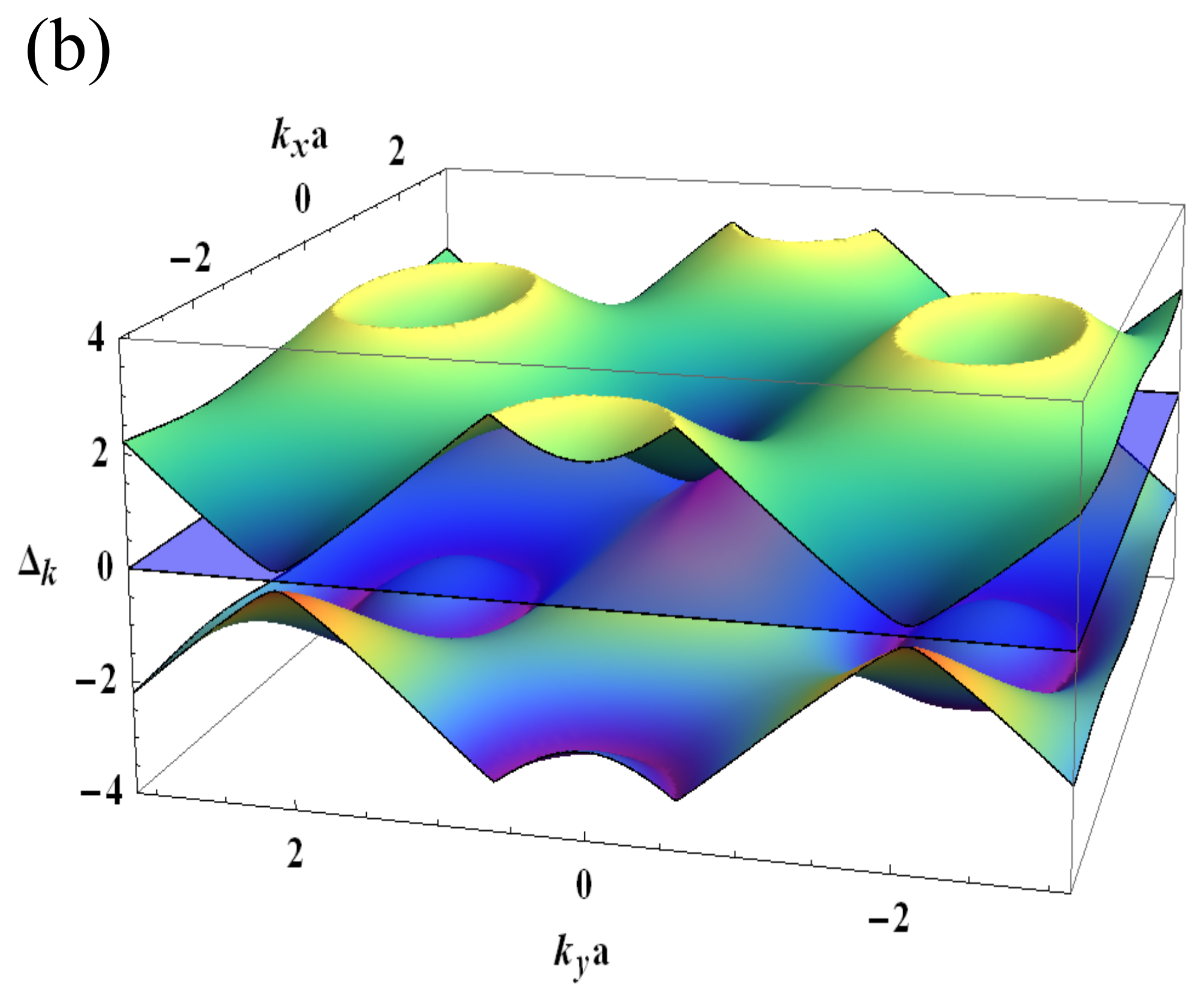}
 \end{minipage}

\caption{Sketch of the quasienergy spectrum corresponding to periodic kicking along $x$-direction for
(a) $\lambda_x=\gamma T$ and (b) $\lambda_x=3\gamma T$. For, $\lambda_x=\gamma T$,
the spectrum exhibits a gapless line at $k_x=\pi/(\sqrt{3}a)$ along $k_y$-direction
and a linear dispersion along $k_x$-direction around the gapless line as described
in Eq.(\ref{alphaX1}). For $\lambda_x=3\gamma T$, the spectrum exhibits semi-Dirac dispersion,
linear along $k_y$ and quadratic along $k_x$ as obtained in Eq.(\ref{alphaX3}).}
\label{fig:alphax_kick}
\end{figure}

The velocity along $y$-direction vanishes for $\lambda_x=\gamma T$. This feature indicates that the quasienergy spectrum
around the gapless point should be linear along $k_x$-direction. Let us understand this argument quantitatively.
In this particular case,
the gapless conditions reduce to $\sin(3ak_y^0/2)=0$ and $\cos(\sqrt{3}ak_x^0/2)=0$ which 
give gapless point $(k_x^0,k_y^0)=(\pi/(\sqrt{3}a),0)$. We find that near $(k_x^0,k_y^0)$ the quasienergy spectrum 
behaves as
\begin{eqnarray}\label{alphaX1}
&&\Delta_{k}^x=\gamma(\sqrt{3}k_xa-\pi-2),~~{\rm for}~~ k_y=0,\nonumber\\
&&\Delta_{k}^x=0,~~~~~~~~~~~~~~~~~~~~~~~~{\rm for}~~ k_x=\frac{\pi}{\sqrt{3}a}.
\end{eqnarray} 
Therefore, the quasienergy spectrum becomes gapless along the line $k_x=\pi/(\sqrt{3}a)$. In Fig.~\ref{fig:alphax_kick}(a), we plot the 
exact quasienergy spectrum [Eq.(\ref{XKick})] for 
$\lambda_x=\gamma T$. This figure also supports our arguments given above.

There exists another interesting case corresponding to 
$\lambda_x=3\gamma T$. Here, the gapless conditions will be 
determined by $\sin(3ak_y^0/2)=0$ and $\cos(\sqrt{3}ak_x^0/2)=1$ which implies the gapless point $(k_x^0,k_y^0)=(0,0)$. This 
is known as merging of Dirac points, achieved by the application of periodic $\delta$-kicking. Similar phenomenon has been revealed 
in the case of graphene[\onlinecite{Graphn_band, FTP1}] also.
Near $(k_x^0,k_y^0)$ the quasienergy spectrum exhibits the following features
\begin{eqnarray}\label{alphaX3}
&&\Delta_k^x=\frac{2a}{T}\sin\Big(\frac{3\gamma T}{2}\Big)k_y,~~~~{\rm for}~~~k_x=0,\nonumber\\
&&\Delta_{k}^x=\frac{3}{4}\gamma a^2k_x^2,~~~~~~~~~~~~~~{\rm for}~~~k_y=0.
\end{eqnarray}
Thus the quasienergy spectrum exhibits semi-Dirac dispersion which is Dirac like along $k_y$ and quadratic along $k_x$. This 
interesting feature is also depicted in Fig.~\ref{fig:alphax_kick}(b) which is obtained by plotting the 
exact quasienergy spectrum [Eq.(\ref{XKick})] for $\lambda_x=3\gamma T$. 

\subsection{Y-Kicking}
The periodic kicking along the $y$-direction is described by $\lambda_x=0$, $\lambda_y\neq0$, and $\lambda_z=0$.
The Floquet operator in this case is reduced to $U_F^y(T)=e^{-i\lambda_y S_y} e^{-iH_{\bm k}T}$.The corresponding quasienergies are obtained as 
\begin{eqnarray}\label{QE_YKick}
\Delta_{\bm k}^{y0}=0,~~~
\Delta^{y\pm}_{\bm k}=\pm \Delta_{\bm k}^y=\pm\frac{1}{T} \cos^{-1}(\kappa^y),
\end{eqnarray}
where
\begin{eqnarray}
 \kappa^y&=& \frac{1}{2}\Big(\cos\lambda_y-1\Big)\cos^2\theta_{\bm k}
 +\sin\lambda_y \sin\theta_{\bm k}\sin(\omega_{\bm k}T)\nonumber\\
 &+&\frac{1}{2}\Big[\cos^2\theta_{\bm k}+\cos\lambda_y\big(1+\sin^2\theta_{\bm k}\big)\Big]\cos(\omega_{\bm k}T).
\end{eqnarray}

Let us now find the ${\bm k}$-points at which the quasienergy spectrum becomes gapless.
In order to have $\Delta_{\bm k}^y=0$, we need $\kappa^y=1$. This criterion will be fulfilled by the 
following choices: $\cos\theta_{\bm k}=0$, $\sin\theta_{\bm k}=1$, and $\lambda_y=\omega_{\bm k}T$. Therefore, 
at the gapless points $\xi_{\bm k}$ vanishes. This leads to infer that the following conditions would be  
satisfied by the gapless point $(k_x^0, k_y^0)$ simultaneously:
\begin{eqnarray}
&&\cos\Bigg(\frac{\sqrt{3}a k_x^{0}}{2}\Bigg)=\frac{\sqrt{\lambda_y^2+\gamma^2T^2}}{2\gamma T},\nonumber\\
&&\sin\Bigg(\frac{3a k_y^{0}}{2}\Bigg)=\frac{\lambda_y}{\sqrt{\lambda_y^2+\gamma^2T^2}},\nonumber\\
&&\cos\Bigg(\frac{3a k_y^{0}}{2}\Bigg)=-\frac{\gamma T}{\sqrt{\lambda_y^2+\gamma^2T^2}}.
\end{eqnarray}
In this case the upper bound of $\lambda_y$ is
$\lambda_y\leq\sqrt{3}\gamma T$.

We now proceed to calculate effective quasienergy spectrum in the vicinity of the gapless points. Here, we consider 
the following approximations: $\sin\theta_{\bm k}\approx [1-\xi_{\bm k}^2/(2\delta_{\bm k}^2)]$ and 
$\cos\theta_{\bm k}\approx -\xi_{\bm k}/\delta_{\bm k}$. The equation for quasienergy takes the form 
\begin{eqnarray}
\cos\big(\Delta_{\bm k}^yT\big)&=&\Bigg(1-\frac{\xi_{\bm k}^2}{2\delta_{\bm k}^2}\Bigg)\cos(\omega_{\bm k}T-\lambda_y)\nonumber\\
&+&\frac{\xi_{\bm k}^2}{2\delta_{\bm k}^2}\big[\cos(\omega_{\bm k}T)+(\cos\lambda_y-1)\big].
\end{eqnarray}

\begin{figure}[h!]
\centering
 \includegraphics[width=6.5cm, height=5.5cm]{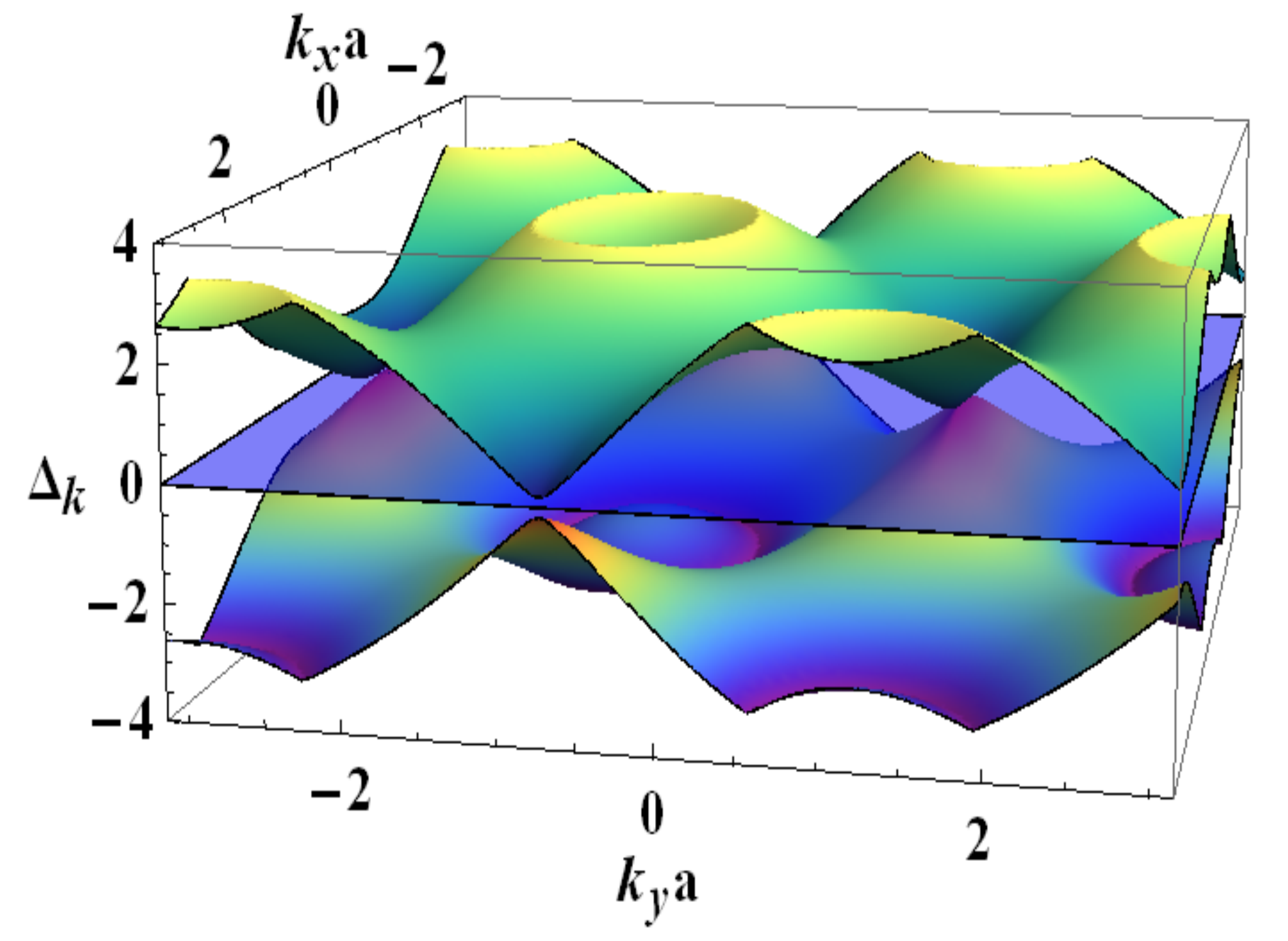}

\caption{Sketch of the exact quasienergy spectrum[Eq.(\ref{QE_YKick})] corresponding
to periodic kicking along $y$-direction for 
$\lambda_y=\sqrt{3}\gamma T$.
Here, the quasienergy spectrum exhibits semi-Dirac dispersion around the gapless point $(0,4\pi/9a)$ as mentioned in Eq.(\ref{YKick}).}
\label{fig:alphay_kick}
\end{figure}

Expanding the cosine functions near the gapless point, we get 
\begin{eqnarray}
&&\big(\Delta_{\bm k}^yT\big)^2=(\omega_{\bm k}T-\lambda_y)^2+\frac{2\xi_{\bm k}^2}{\delta_{\bm k}^2}(1-\cos\lambda_y).
\end{eqnarray}
By considering $\delta_{\bm k}=\lambda_y/(\gamma T)$ at the gapless points, we obtain the 
velocity components of the quasiparticle as 
\begin{eqnarray}
&&v_x=\frac{\sqrt{3}a}{2T}\Bigg[\frac{2\gamma T}{\lambda_y}\sin\Big(\frac{\lambda_y}{2}\Big)-\lambda_y\Bigg]
\sqrt{\frac{3\gamma^2T^2-\lambda_y^2}{\gamma^2T^2+\lambda_y^2}},\nonumber\\
&&v_y=-\frac{3a}{2T}\Big[\gamma T+2\sin\Big(\frac{\lambda_y}{2}\Big)\Big].
\end{eqnarray}
For $\lambda_y=\sqrt{3}\gamma T$, $v_x$ vanishes. Therefore, we argue that the quasienergy spectrum would be linear along $k_y$.
In support of our argument we are going to see the behavior of quasienergy about the gapless point corresponding to
$\lambda_y=\sqrt{3}\gamma T$. In this case, we find the gapless point $(k_x^0, k_y^0)=(0,4\pi/(9a))$.
Around this point, the quasienergy spectrum exhibits following semi-Dirac feature:
\begin{eqnarray}\label{YKick}
&&\Delta_k^y(k_x^0,k_y)=\frac{3a\bar{k}_y}{2T}\sqrt{\gamma^2T^2+4\sin^2\Big(\frac{\sqrt{3}\gamma T}{2}\Big)},\nonumber\\
&&\Delta_k^y(k_x,k_y^0)=\frac{\sqrt{3}k_x^2a^2}{8T}\sqrt{9\gamma^2T^2+4\sin^2\Big(\frac{\sqrt{3}\gamma T}{2}\Big)},
\end{eqnarray}
where $\bar{k}_y=k_y-k_y^0$. These features are 
clearly depicted in Fig.~\ref{fig:alphay_kick}.

\subsection{Z-Kicking}

For z-kicking i.e. $\lambda_x=0$, $\lambda_y=0$, and 
$\lambda_z\neq0$, the Floquet operator becomes $U_F^z(T)=e^{-i\lambda_z S_z} e^{-iH_{\bm k}T}$. In this case we find 
\begin{eqnarray}\label{QE_ZKick}
\Delta_{\bm k}^{z0}=0,~~~
\Delta^{z\pm}_{\bm k}=\pm \Delta_{\bm k}^z=\pm\frac{1}{T} \cos^{-1}\big(\kappa^z\big),
\end{eqnarray}
where $\kappa^z=\cos^2\big(\lambda_z/2\big)\cos\big(\omega_{\bm k}T\big)-\sin^2\big(\lambda_z/2\big)$.

\begin{figure}[h!]
 \begin{minipage}[b]{\linewidth}
\centering
 \includegraphics[width=6.5cm, height=4.5cm]{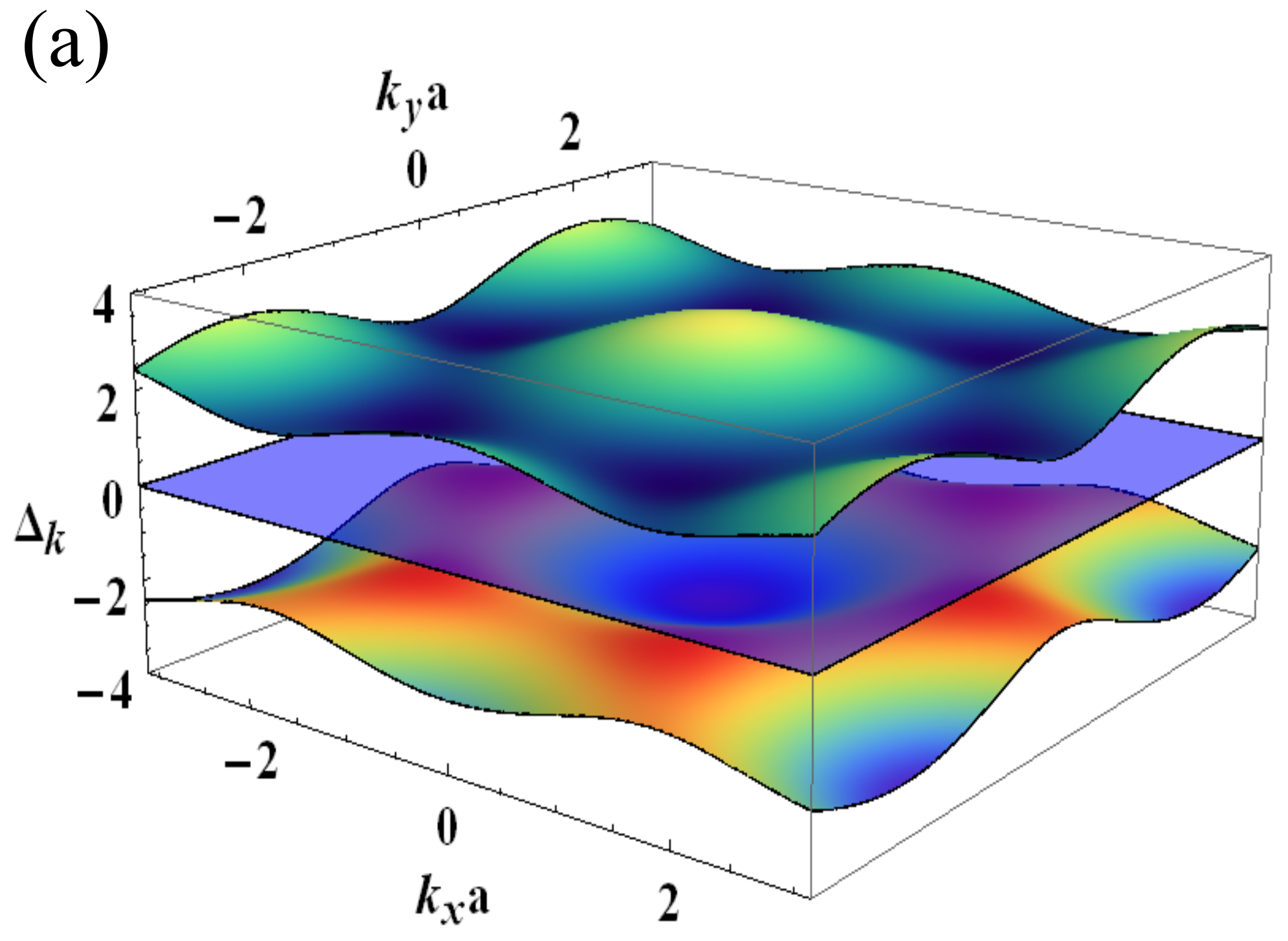}
 \end{minipage}
\vspace{0.1em}

\begin{minipage}[b]{\linewidth}
 \centering
 \includegraphics[width=6.5cm, height=4.5cm]{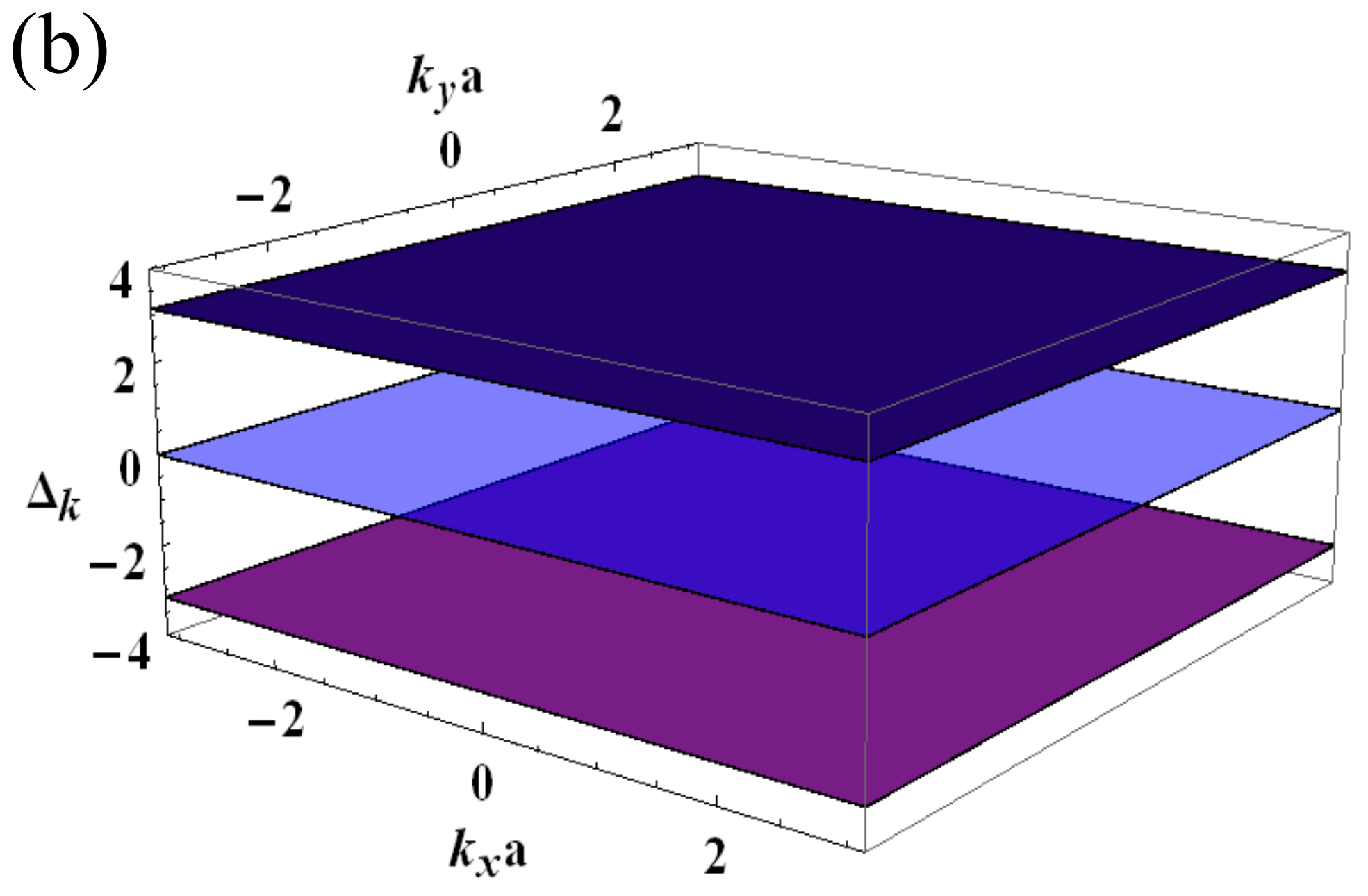}
 \end{minipage}
\caption{Sketch of the exact quasienergy spectrum corresponding to the periodic kicking along the $z$-direction for (a)
$\lambda_z=\pi/2$ and (b) $\lambda_z=\pi$. The spectrum is gapped for each value of $\lambda_z$. Completely flat quasienergy bands 
are obtained for $\lambda_z=\pi$.}
\label{fig:alphaz_kick}
\end{figure}

Let us check whether gapless points in this particular case still exist or not. If the quasienergy spectrum possesses any 
gapless point we must have $\kappa^z=1$ which implies $\cos(\omega_{\bm k}T)=1+2\tan^2(\lambda_z/2)$. But this cannot be true because $\cos(\omega_{\bm k}T)$ is bounded between $-1$ and $1$. Therefore, the quasienergy spectrum is gapped everywhere.
Consider the Dirac points of the unperturbed system where the conduction band touches the valence band.
Here, $\omega_{\bm k}=0$, which consequently implies $\Delta_{\bm k}=\lambda_z/T$. A gap is opened up at the Dirac points. We have $\Delta_{\bm k}=\pi/T$ for $\lambda_z=\pi$. 
Therefore, the quasienergy spectrum is independent of the wave vector. This feature is shown in Fig. 4. 
Similar feature is also obtained in the case of graphene[\onlinecite{Graphn_band}]. 
At ${\bm k}$-values other than the Dirac points i.e.  $\omega_{\bm k}\neq0$, we also have $\Delta_{\bm k}=\pi/T$ for $\lambda_z=\pi$.
This absolutely flat quasienergy bands lead to dynamical localization of wave packet. On the other hand, for 
$\lambda_z=\pi/2$, $\kappa^z=\big[\cos\big(\omega_{\bm k}T\big)-1\big]/2$. Therefore, $\Delta^{z\pm}_{\bm k}$ become dispersive unlike the case for $\lambda_z=\pi$.
The dispersive and flat quasienergy bands are depicted in Fig.~\ref{fig:alphaz_kick}(a) and (b) for $\lambda_z=\pi/2$ and $\pi$, respectively, corroborating the above analytical findings.  

\textcolor{black}{The particle-hole symmetry is preserved for the driven Dice model  irrespective of the direction of the kick. This is clearly evident from the structure of quasi-energy dispersions $\Delta_{\bm k}^{i0}=0$ and $\Delta^{i+}_{\bm k}= -\Delta^{i-}_{\bm k}$
with
$\Delta^{i\pm}_{\bm k}=\pm \Delta_{\bm k}^i=\pm\frac{1}{T} \cos^{-1}\big(\kappa^i\big)$ for the kick along $i$  
direction with $i=x,~y,~z$. As stated above, the particle-hole symmetry demands
 ${\mathcal P} U_F^{i}({\bm k},T) {\mathcal P}^{-1}= U_F^{i}(-{\bm k},T)$ referring  to the fact that $\Delta_{\bm k}^{i\pm} =- \Delta_{\bm k}^{i \mp} $ and  ${\mathcal P} |\phi_{\pm}^i(\bm k)\rangle = |\phi_{\mp}^i(-\bm k)\rangle$. The detailed calculation is provided in the Appendix where  $U_F^{x}({\bm k},T)$ is computed. Furthermore, for $\alpha=0$ (graphene case), one can find quasi-energies to be  paticle-hole symmetric  referring to the fact driven system preserves particle-hole symmetry [\onlinecite{Graphn_band}]. 
Therefore, the periodically kicked  Dice model and graphene behave identically as far as the particle-hole symmetry is concerned.}

\section{Wave Packet Dynamics in a Kicked Dice Lattice}

Having discussed the Floquet quasienergy dispersion, we now numerically study the time evolution of a wave packet on the dice lattice. 
Consider an initial Gaussian wave packet $\Psi$ in two spatial dimensions, with an initial 
momentum $\bm k_i=(k_{ix},k_{iy})$ and a width
$\sigma$ as given below
\begin{equation}
 \Psi(\bm{r},t=0)=\frac{1}{\sqrt{2\pi\sigma^2}} \exp(-\frac{r^2}{4 \sigma^2}) ~
\exp(i \bm k_i \cdot \bm{r}),
\end{equation}
which is normalized such that $\int d\bm{r}\, \vert \Psi \vert^2=1$.
The wave packet is centered around $\bm {r}=(0,0)$ and decaying in $x$ and $y$ direction
in a uniform (circularly symmetric) manner  with localization length $2\sigma$. One can consider the Fourier transform of $\Psi$ to obtain the ${\bm k}$ resolved wave function 
\begin{equation} \Psi(\bm k,t=0)= \sqrt{8 \pi \sigma^2} \exp[- \sigma^2 \{ (k_x-k_{ix})^2 + 
(k_y-k_{iy} )^2 \}]
\label{eq:ini_wp}
\end{equation}
such that $\frac{1}{(2\pi)^2}\int dk_x dk_y \vert \Psi({\bm k})\vert^2 = 1$. We use the momentum space
wave function [Eq.(\ref{eq:ini_wp})] to study the stroboscopic dynamics at integer multiple of $T$. Our aim is to understand the subsequent wave packet dynamics from the quasienergy dispersion 
$\Delta_{\bm k}$.

The wave packet after $n$ kicks i.e. at time $t=nT$ will be obtained by $n$-times kick-to-kick
operator $U_F(nT)$ onto $\Psi({\bm k},0)$. One has to write the initial wave packet 
in the basis of the dice lattice Hamiltonian [Eq.(\ref{Ham_alpha}) for $\alpha=1$] as 
$ \Psi({\bm k},0) \to  \Psi({\bm k},0) (0,1,-i)^{\mathcal T}/\sqrt{2}$ with ${\mathcal T}$ being the transpose.
The static dice model shows a zero energy flat band and two dispersive valence 
and conduction bands at finite energies. In the absence of any kicking, the negative energy
valence and zero energy flat bands are occupied. 
The wave packet movement depends on these initially occupied bands. 
Upon periodic kicking,
therefore, we have 
\begin{eqnarray}
 \Psi({\bm k},t=nT)=U_F(\bm k, nT)\Psi({\bm k},0)=\Big[U_F(\bm k, T)\Big]^n \Psi({\bm k},0).\nonumber\\
\end{eqnarray}
The driven system continues to preserve the translational symmetry that allows us to study each
momentum mode ${\bm k}=(k_x,k_y)$ separately. Thus obtained the ${\bm k}$ resolved 
wave function  $ \Psi({\bm k},t=nT)$ is Fourier transformed using the real space lattice 
structure of dice lattice model. For this, one needs the real space position of each site on the
dice lattice ${\bm r}=(x,y)$. We consider the momentum space Brillouin zone as the rhombus with the
vortices $(\pm 2\pi/\sqrt{3},0)$ and  $(0,\pm 2\pi/3)$ and center lying at $(0,0)$. 
We consider $N=40$ blocks of graphene and the total number of lattice sites is $3N(2N+1)/2=4860$.
We take periodic boundary condition in real and momentum space to compute the wave packet dynamics.


\begin{figure}[h!]
\centering
 \includegraphics[width=7.0cm, height=5.5cm]{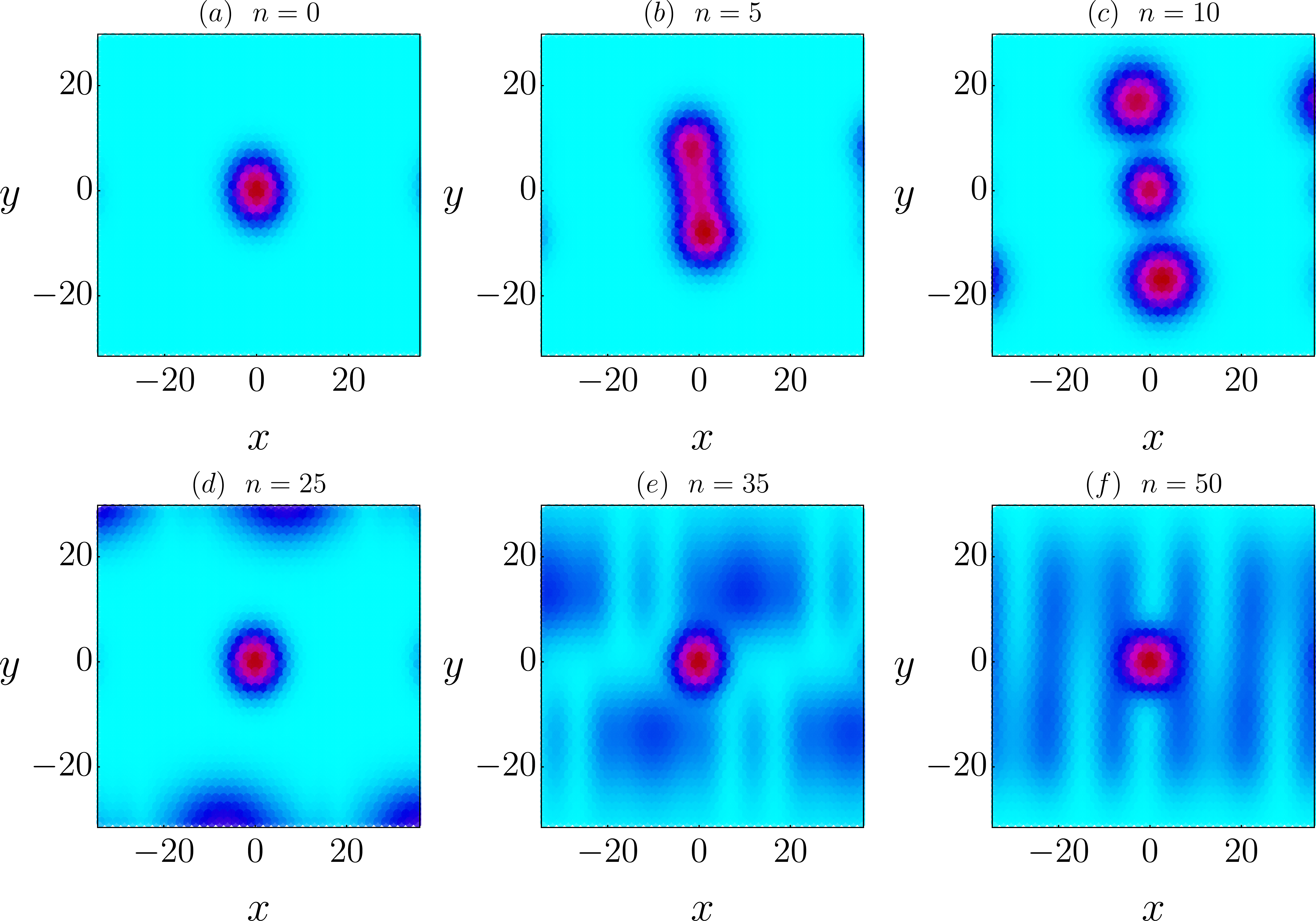}
 \includegraphics[width=1.10cm, height=3.3cm]{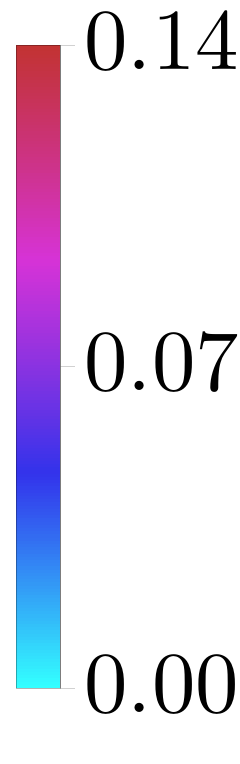}
\caption{We show the time evolution of the 
initial wave packet centered at ${\bm r}=(0,0)$ at $t=0$ in (a), 
$t=5T$ in (b), $t=10T$ in (c), $t=25T$ in (d), $t=35T$ in (e), and
$t=50T$ in (f) for $\lambda_x=\gamma T$, $\lambda_y=\lambda_z=0$ with
$k_{ix} a=0.0$ and $k_{iy} a=1.0 $ and $\sigma=5a/\sqrt{2}$. 
One can clearly observe that the wave packet evolves with a net 
velocity along $y$-direction; however, with time the wave packet spreads in $x$-direction also. At later time $t=50T$ as a result, the localized fringe
structure along $y$-direction is observed. The important point to note here 
is that there always exists a substantially localized wave packet centered
at  ${\bm r}=(0,0)$ irrespective of the kick as there always exists the flat band in the quasienergy dispersion.}
\label{figure:LDOS_alphax_1p0}
\end{figure}


We first study the wave packet dynamics with 
$x$-kicking i.e., $\lambda_x=\gamma T$, and $\lambda_y=\lambda_z=0$. We here show that the wave packet spreads along
$y$-direction (see Fig.~\ref{figure:LDOS_alphax_1p0}(a,b,c) ). This is due to the fact that the initial non-zero momentum 
is chosen only along $y$-direction:  $k_{ix}a =0.0$ and $k_{iy}a =1.0 $. 
With increasing time, fringes like localized structure are formed (see Fig.~\ref{figure:LDOS_alphax_1p0}(d,e,f)).
The time evolved wave packet thus spreads all over the 2D dice lattice. It is interesting to note that there
always exists a wave packet centered around ${\bm r}=(0,0)$ throughout the time evolution. 
The intermediate energy band continues to remain flat even under 
the Floquet dynamics as shown in Fig.~\ref{fig:alphax_kick}. As a result,
the quasivelocity becomes zero for this band while for the other two finite energy bands, the quasivelocities
become finite and opposite in sign. Since the initial wave packet is projected 
in these states, we always get the signature of flat band in the centrally localized wave packet. Due to the
finite quasivelocities of the dispersive bands, the wave packet spreads in both the direction as the initial 
momentum  $k_{ix} a=0.0$ and $k_{iy} a=1.0 $ for 
wave packet is not located over the dispersionless line with $k_x=\pi/\sqrt{3} a$.

\begin{figure}[h!]
\centering
 \includegraphics[width=7.0cm, height=5.5cm]{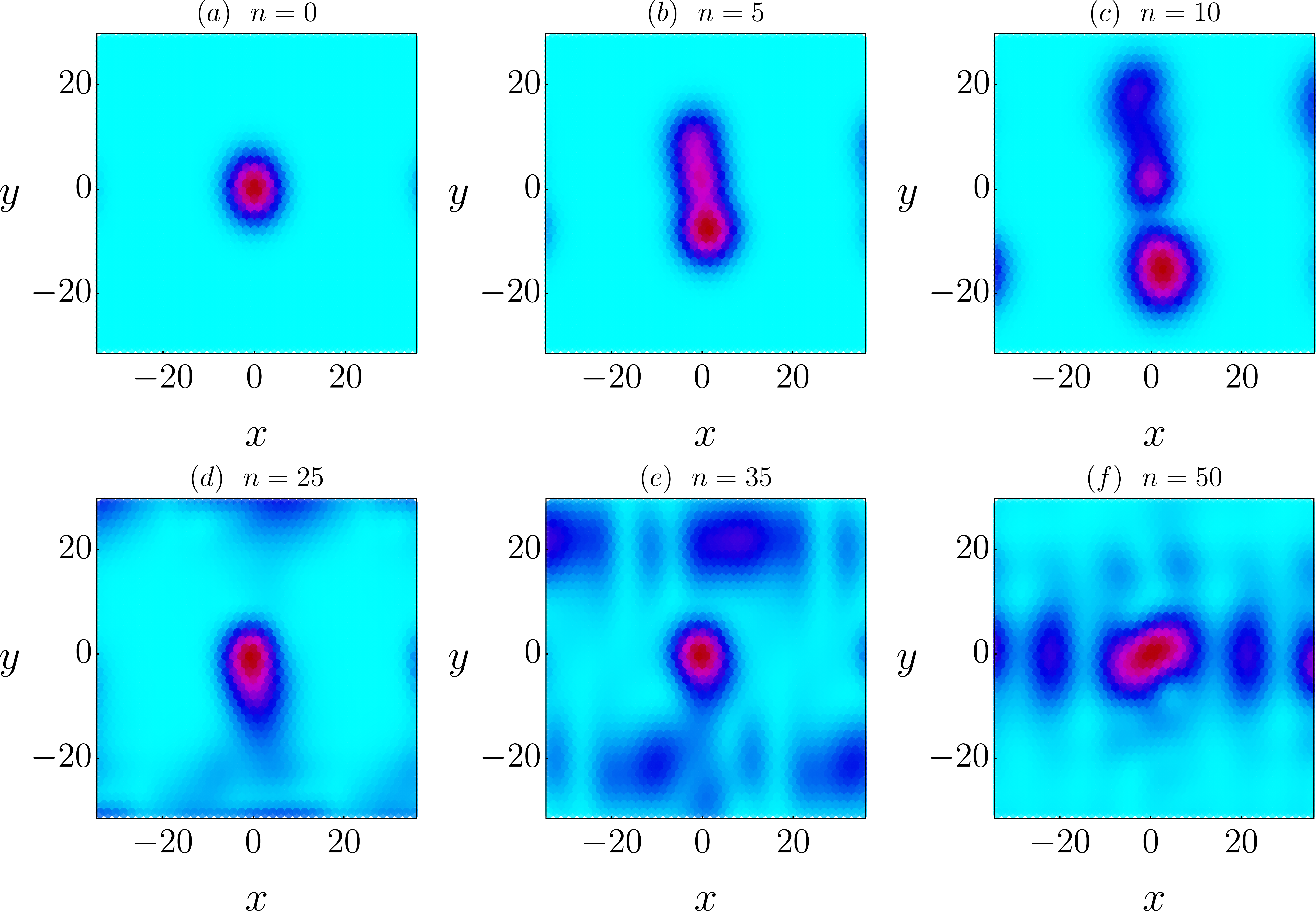}
\caption{We repeat Fig.~\ref{figure:LDOS_alphax_1p0} with $\lambda_x=3\gamma T$. The wave packet moves
along $y$-direction, however, the weight transfer in a non-uniform manner. The wave packet moving in 
the $-y$-direction shows higher degree of localization as compared to the wave packet moving in $+y$-direction. 
At later time, the localized fringe structure along $y$-direction, as observed for 
Fig.~\ref{figure:LDOS_alphax_1p0}, becomes distorted. The localized structure of
wave packet at  ${\bm r}=(0,0)$ remains unaltered as the flat band continues to exist in the 
dynamics.}
\label{figure:LDOS_alphax_3p0}
\end{figure}

\begin{figure}[h!]
\centering
 \includegraphics[width=7.0cm, height=5.5cm]{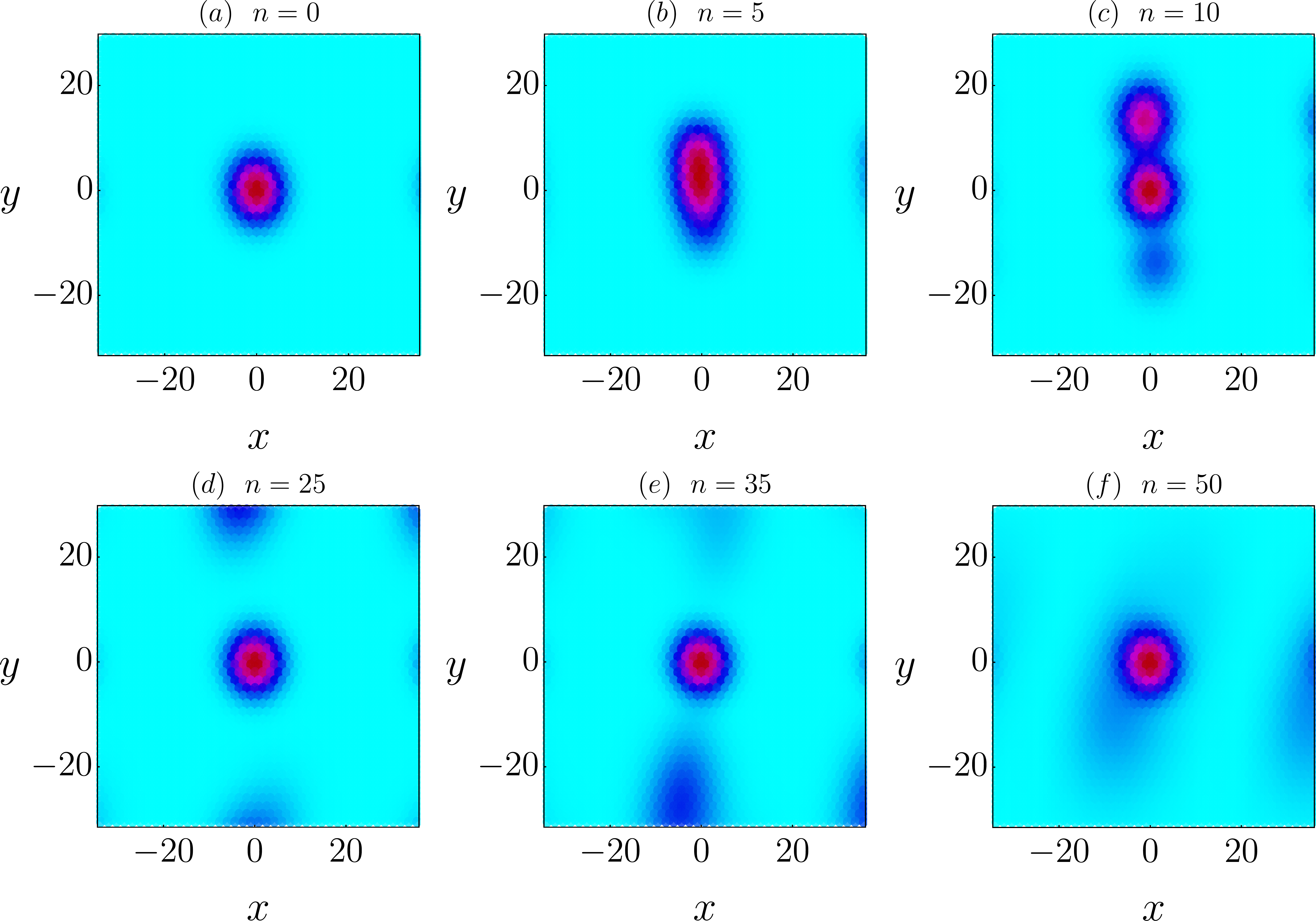}
\caption{We repeat Fig.~\ref{figure:LDOS_alphax_1p0} with $\lambda_y=\sqrt{3}\gamma T$, and  $\lambda_x=\lambda_z=0$.
The wave packet moves substantially along 
$+y$-direction. At later time, the fringe like structure of localized
wave packet is not observed unlike the 
$x$-kicking. The flat band continues to persist as the  centrally localized
wave packet remain there with time.}
\label{figure:LDOS_alphay_sr3p0}
\end{figure}

We note that $v_x$ can  become finite with $v_y =0$ for $\bm k_i$ chosen over the dispersionless line causing the initial wave packet to move along $x$-direction only. 
At a later time, the  interference between different momenta leads to fringe like structure in real space. 
For $\lambda_x=3\gamma T$, the Floquet quasienergy dispersion changes leading to a  different quasivelocity
as compared to the previous case with $\lambda_x=\gamma T$.  For small time,  wave packet disperses along $-y$-direction more strongly
than $+y$-direction (see Fig.~\ref{figure:LDOS_alphax_3p0} (a,b,c)). 
The wave packet spreading at later times becomes different and fringe pattern gets
distorted (see Fig.~\ref{figure:LDOS_alphax_3p0} (d,e,f)). A comparison between Fig.~\ref{figure:LDOS_alphax_1p0} and Fig.~\ref{figure:LDOS_alphax_3p0} clearly suggests that the
wave packet movement changes once there exists a
gapless line in the quasienergy dispersion for $\lambda_x=\gamma T$. 
In a similar spirit,
we study the wave packet dynamics for $\lambda_y=\sqrt{3}\gamma T$ as shown in Fig.~\ref{figure:LDOS_alphay_sr3p0}. 
The wave packet dynamics shows quantitatively different behavior with respect to $x$-kicking however, qualitatively 
$x$- and $y$-kicking result in similar behavior. The delocalization of the Gaussian wave packet is observed.

\begin{figure}[h!]
\centering
 \includegraphics[width=7.0cm, height=5.5cm]{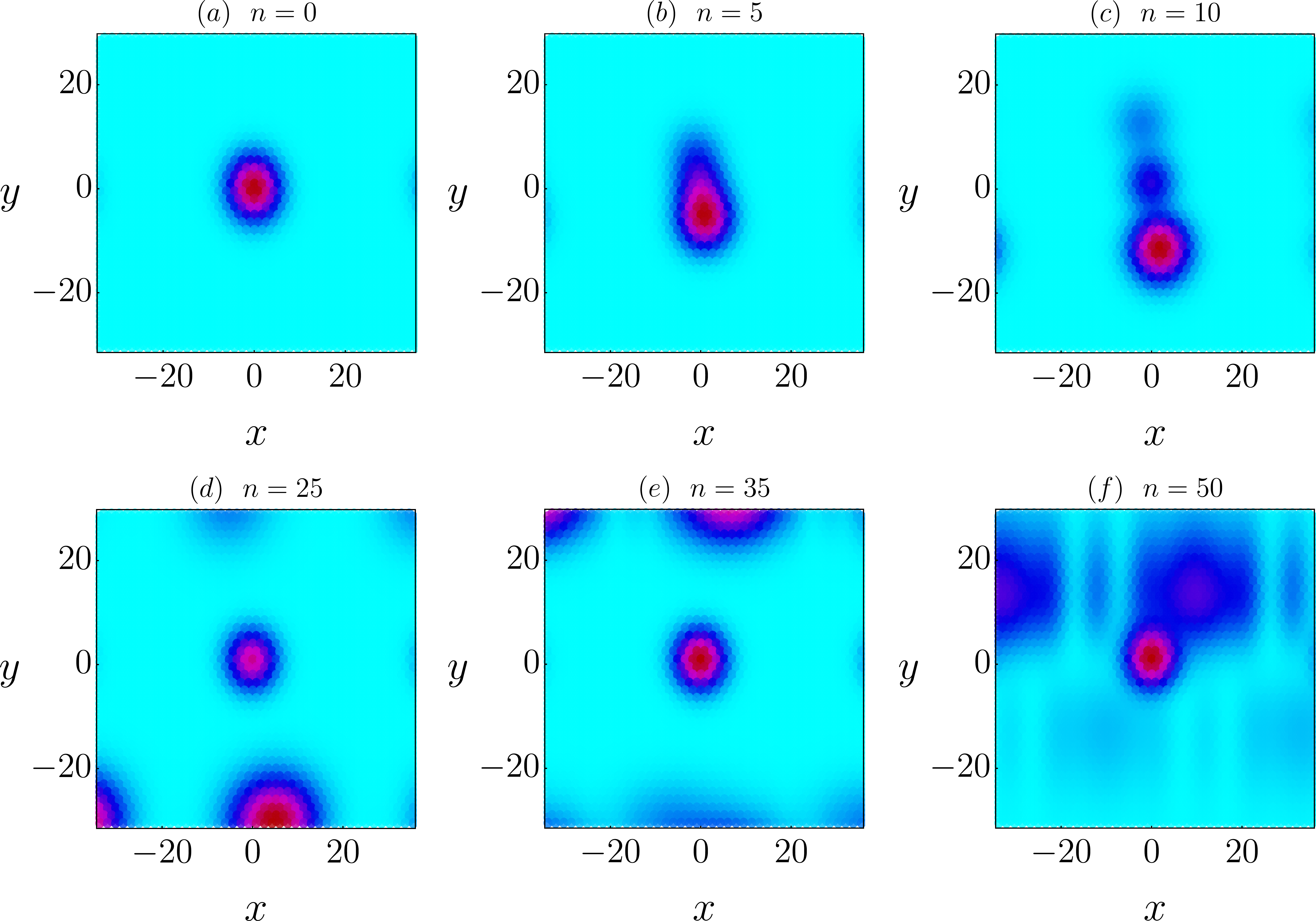}
\caption{We repeat Fig.~\ref{figure:LDOS_alphax_1p0} with $\lambda_z=\pi/2$, and  $\lambda_x=\lambda_y=0$.
For small time, the wave packet as a whole moves along $-y$-direction. At later time, the localized fringe structure
arises in addition to the centrally localized wave packet.}
\label{figure:LDOS_alphaz_pib2}
\end{figure}

\begin{figure}[h!]
\centering
 \includegraphics[width=7.0cm, height=5.5cm]{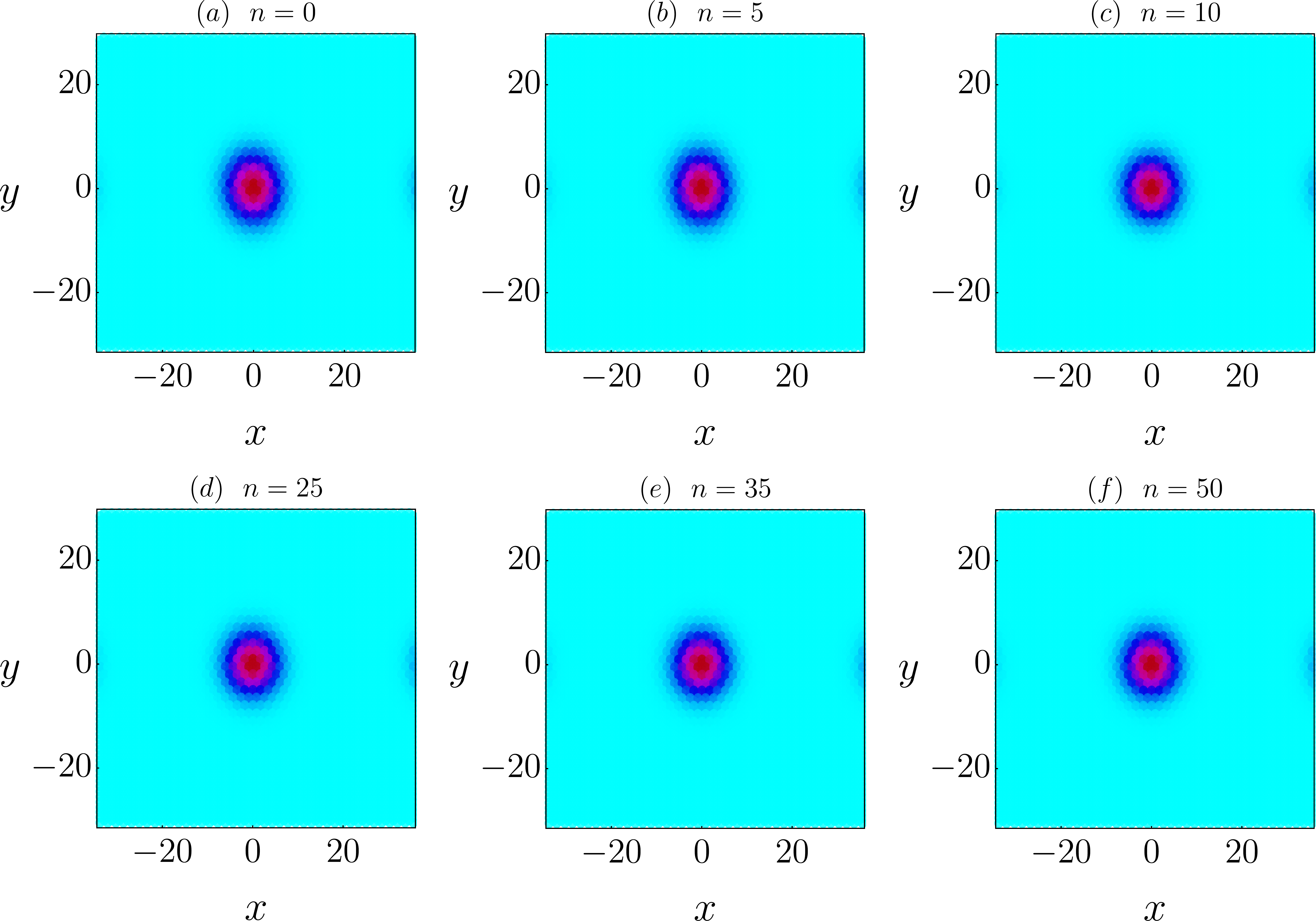}
\caption{We repeat Fig.~\ref{figure:LDOS_alphaz_pib2} with
$\lambda_z=\pi$. Here, we clearly observe the dynamical localization as the initial wave packet remains
frozen at its localization core. This situation is corroborated with the fact that all three quasibands become 
flat and hence the quasivelocity vanishes identically for all these bands. This is in stark contrast to all of
the previous situation where initial wave packet disperses with time.}
\label{figure:LDOS_alphaz_pi}
\end{figure}

We shall now analyze the $z$-kicking as shown in Fig.~\ref{figure:LDOS_alphaz_pib2} and 
Fig.~\ref{figure:LDOS_alphaz_pi} for $\lambda_z=\pi/2$ and 
$\lambda_z=\pi$, respectively. The stark distinction can be noticed for the later case where
dynamical localization is clearly observed. The initial Gaussian wave packet, centered
around ${\bm r}=(0,0)$, is frozen with time. This can be explained by the fact that all
the Floquet quasienergy bands becomes flat as shown in Fig.~\ref{fig:alphaz_kick}. 
The quasivelocity vanishes for all the bands identically leading to the dynamically 
localized wave packet. On the other hand, with $\lambda_z=\pi/2$, the initial Gaussian
wave packet does not get delocalized for small times rather it moves as a
whole along $-y$-direction (see Fig. \ref{figure:LDOS_alphaz_pib2} (a,b,c)). At later time, 
finite quasivelocity causes the wave packet to spread over the real space lattice  (see Fig. \ref{figure:LDOS_alphaz_pib2} (d,e,f)).    
This can be explained from the fact that gapped quasienergy  (see Fig. \ref{fig:alphaz_kick} (a)) spectrum is actually  dispersive in nature.

\section{Results For a kicked $\alpha$-T$_3$ lattice}
\begin{figure}[h!]
 \begin{minipage}[b]{\linewidth}
\centering
 \includegraphics[width=6.5cm, height=4.5cm]{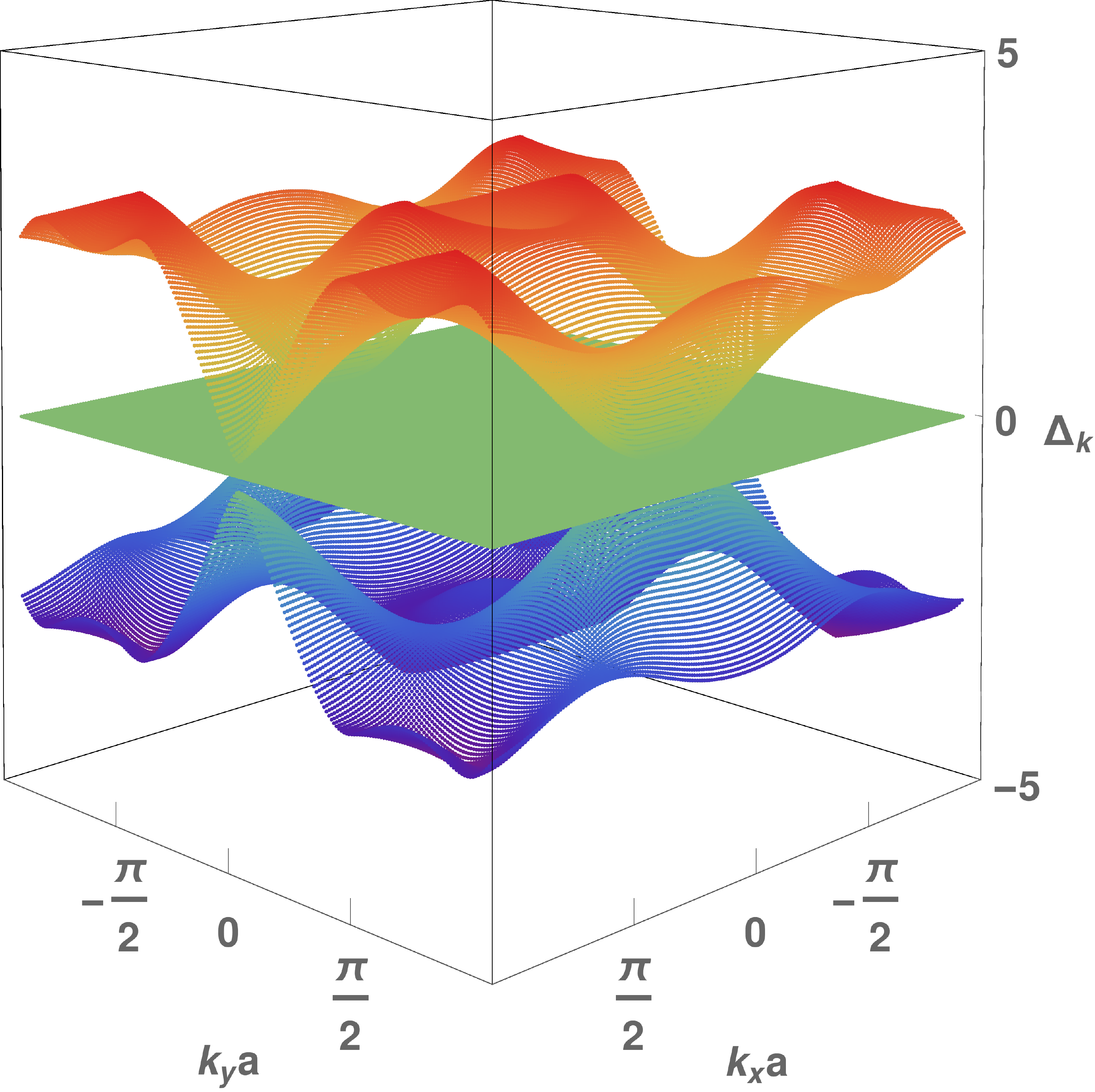}
 \end{minipage}
\vspace{0.1em}

\begin{minipage}[b]{\linewidth}
 \centering
 \includegraphics[width=6.5cm, height=4.5cm]{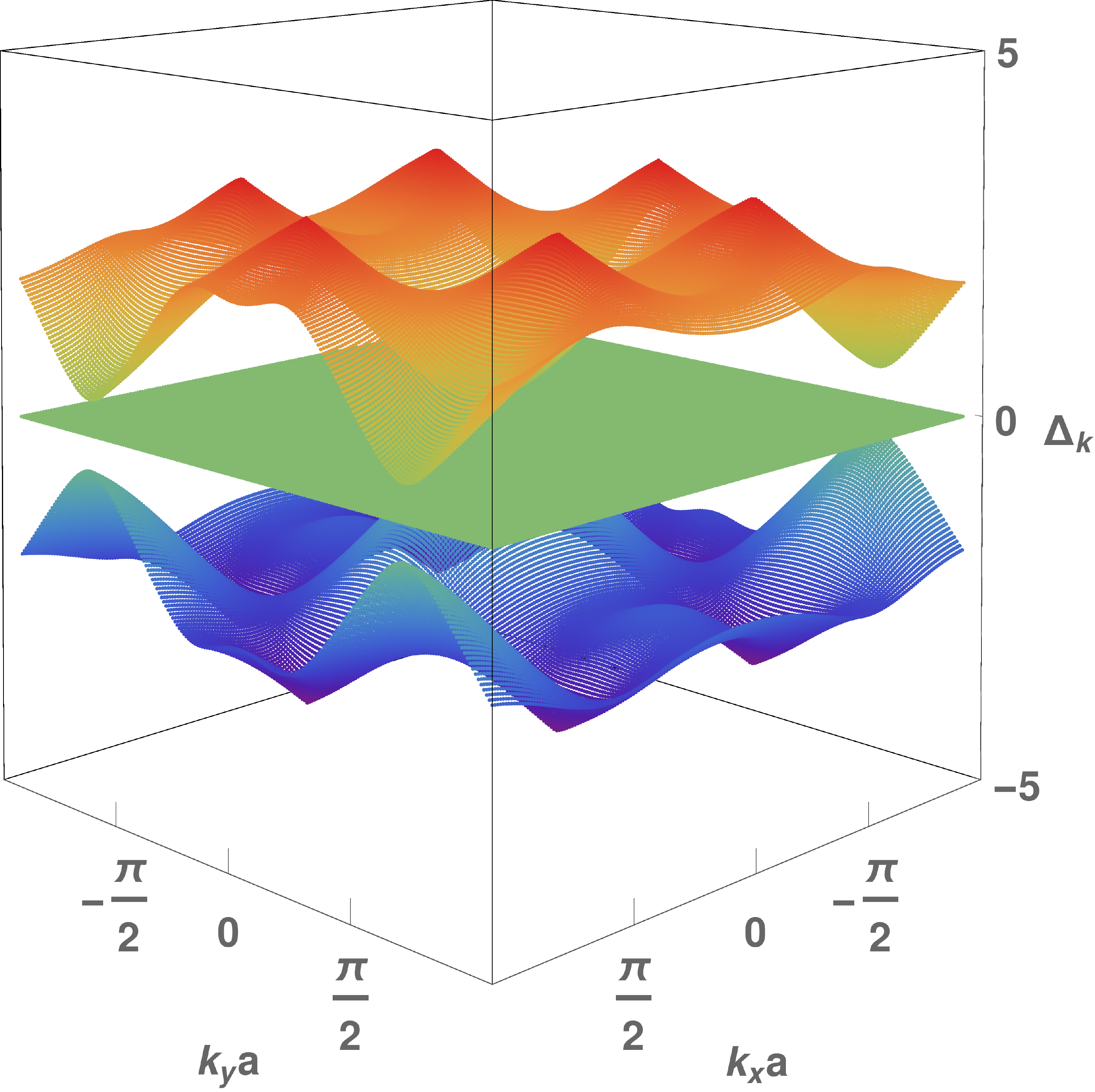}
 \end{minipage}

\caption{\textcolor{black}{Quasienergy dispersion of a periodically kicked $\alpha$-T$_3$ lattice along $x$-direction for $\alpha=0.5$. The upper(lower) panel corresponds to $\lambda_x=\gamma T(3\gamma T)$.}}
\label{fig:alphaT3_Xkick}
\end{figure}

\textcolor{black}{
We now proceed to calculate the quasienergy dispersion of an $\alpha$-T$_3$ lattice corresponding to the periodic kicking in different directions just like the dice lattice case.  We will calculate the quasienergy dispersion numerically as it is difficult to find the quasienergy dispersions in closed forms, particularly in the case of $x$- and $y$-kickings for $0<\alpha<1$. However, some analytical results are possible to obtain in the case $z$-kicking from which some interesting physics can be extracted. In Fig.
\ref{fig:alphaT3_Xkick}, the quasienergy dispersion corresponding to the kicking in $x$-direction for an intermeadiate $\alpha=0.5$ is shown. The upper and lower panel represent two different kicking strength, namely, $\lambda_x=\gamma T$ and
$\lambda_x=3\gamma T$, respectively. Comparing with Fig. \ref{fig:alphax_kick}, we see that the quasienergy dispersion is changed significantly at an intermediate $\alpha$, but the particle-hole symmetry is preserved.} 
\textcolor{black}{A close inspection suggests  
$\Delta_{\bm k}^{i0}=0$ and $\Delta^{i+}_{\bm k}= -\Delta^{i-}_{\bm k}$ with $i=x,~y$ referring to the fact that 
${\mathcal P} U_{F\alpha}^{x,y}({\bm k},T) {\mathcal P}^{-1} = U_{F\alpha}^{x,y}(-{\bm k},T)$. Therefore, 
for any intermediate value $0<\alpha <1$, the kicked $\alpha$-T$_3$ model, with planar kicking protocol, preserves particle-hole symmetry similar to the kicked dice model.   
}

\textcolor{black}{
We have seen that a quasiparticle in a dice lattice exhibits dynamical localization phenomenon 
under periodic kicking in the transverse direction like graphene. Now we would like to address the possibility of occurring  such phenomenon in 
$\alpha$-T$_3$ lattice. Here, we consider the periodic kicking along $z$-direction alone because this particular type of kicking protocol with certain strength was responsible for the dynamical localization of wave packet in the case of dice lattice.
The characteristic equation of the Floquet operator $U_{F\alpha}^z(T)=e^{-i{\lambda}_z S_z^\alpha} e^{-iH_{\bm k}^\alpha T}$ becomes
\begin{eqnarray}\label{chart}
x^3-\beta x^2+\beta^\ast x-1=0,
\end{eqnarray}
where
\begin{eqnarray}
\beta&=&\Omega_{\bm k}e^{2i\lambda_z\cos(2\phi)}+\big(\Omega_{\bm k}\sin^2\phi+\cos^2\phi\big)e^{2i\lambda_z\sin^2\phi}\nonumber\\
&+&\big(\Omega_{\bm k}\cos^2\phi+\sin^2\phi\big)e^{-2i\lambda_z\cos^2\phi}.
\label{beta_general}
\end{eqnarray}
Here, $\Omega_{\bm k}=\cos(\omega_{\bm k}T)$.
The roots of Eq.(\ref{chart}) simply gives $e^{\pm i\Delta_{\bf k}T}$. It is clear from Eq.(\ref{chart}) that $x=1$ 
cannot be a root. Therefore, we argue that the flat quasienergy band with $\Delta_{\bm k}=0$ does not exist in an irradiated 
$\alpha$-T$_3$ lattice. In other words, the periodic $\delta$-kicking breaks the flat band of the $\alpha$-T$_3$ lattice for 
all values of $\alpha$ except $\alpha=0$ and $\alpha=1$. As a consequence the dynamical localization of wave packet under 
periodic driving occurs only in two extreme limits of $\alpha$-T$_3$ lattice, namely $\alpha=0$ (Graphene) and $\alpha=1$ (Dice Lattice).}

\begin{figure}[h!]
 \begin{minipage}[b]{\linewidth}
\centering
 \includegraphics[width=6.5cm, height=4.5cm]{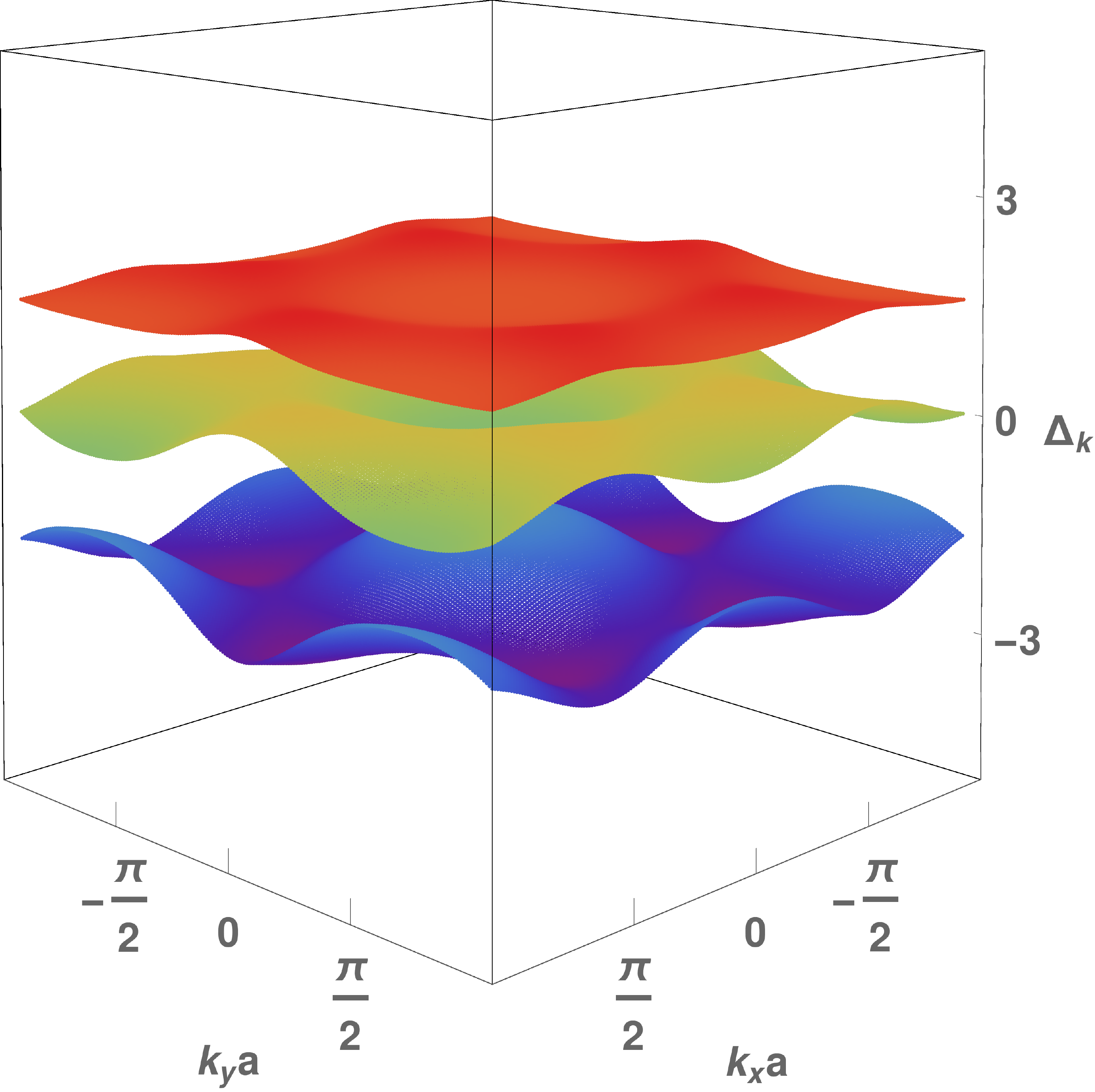}
 \end{minipage}
\vspace{0.1em}

\begin{minipage}[b]{\linewidth}
 \centering
 \includegraphics[width=6.5cm, height=4.5cm]{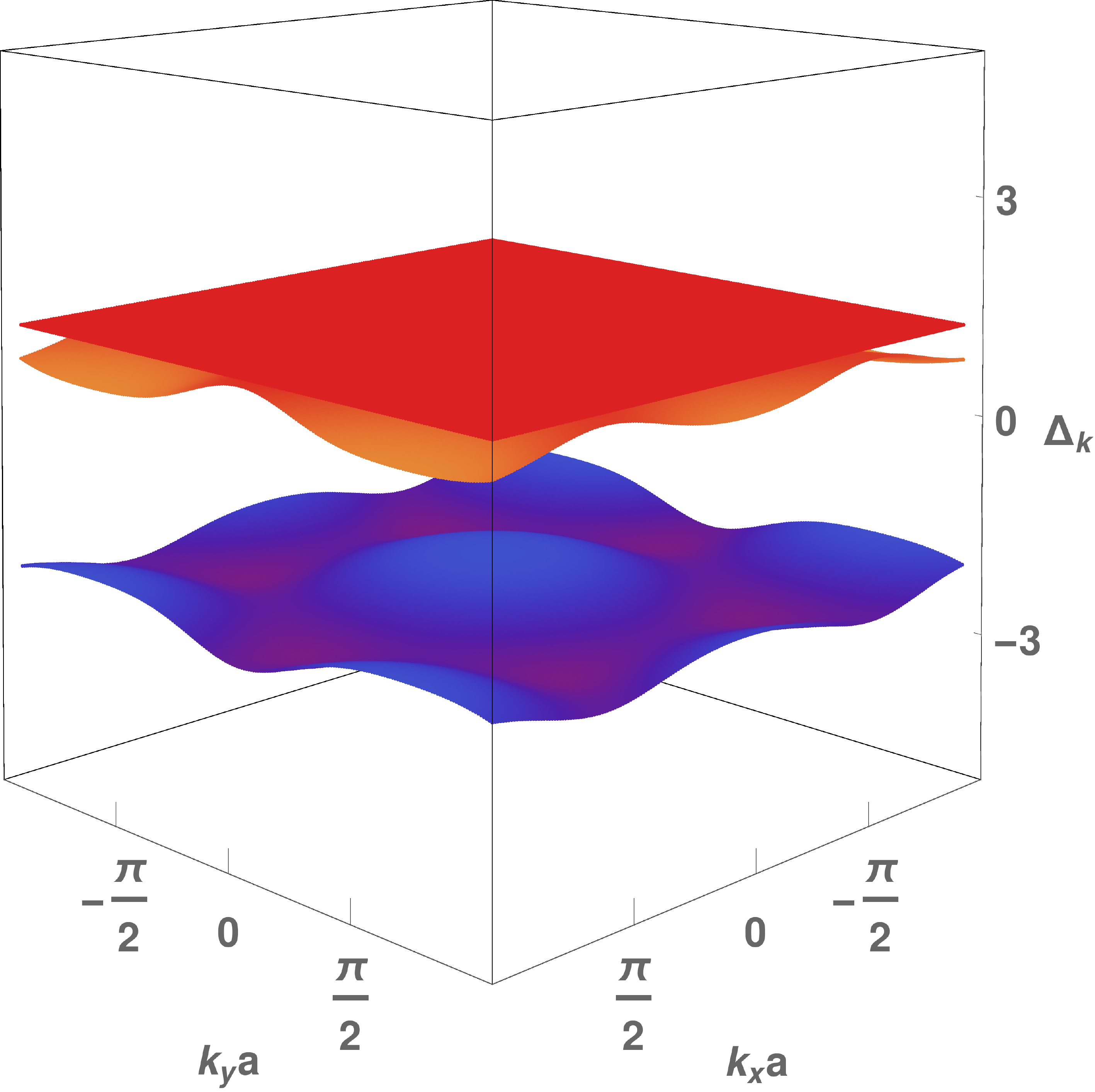}
 \end{minipage}

\caption{\textcolor{black}{Quasienergy dispersion of a periodically kicked $\alpha$-T$_3$ lattice along $z$-direction for $\alpha=0.5$. The upper(lower) panel corresponds to $\lambda_z=(\pi/2)(\pi)$.}}
\label{fig:alphaT3_Zkick}
\end{figure}
\textcolor{black}{
We numerically evaluate the quasienergy spectrum for two different kicking strength $\lambda_z=\pi/2$ and $\pi$ at an intermediate $\alpha=0.5$. This dispersion is depicted in Fig. \ref{fig:alphaT3_Zkick}. Interestingly, the kicking along $z$-direction breaks the particle-hole symmetry for $0<\alpha<1$. This kind of symmetry breaking was addressed earlier when an $\alpha$-T$_3$ lattice is subjected to a circularly polarized radiation propagating in a perpendicular direction[\onlinecite{Sym8}]. We thus conclude that the kicking in perpendicular direction breaks the particle-hole symmetry of $\alpha$-T$_3$ model for $0<\alpha<1$ while the kicking in parallel direction preserves the particle-hole symmetry.}
\textcolor{black}{Unlike the kicked dice model, we find $\Delta_{\bm k}^{z0} \neq 0$ and $\Delta^{z+}_{\bm k}\neq -\Delta^{z-}_{\bm k}$ suggesting the fact that ${\mathcal P} U_{F\alpha}^{z}({\bm k},T) {\mathcal P}^{-1} \neq U_{F\alpha}^{z}(-{\bm k},T)$ (See Appendix for a detail derivation). 
This is in contrast to the kicked dice model, where particle-hole symmetry is alsways respected for perpendicular $z$-kicks as well as planar $x$, $y$-kicks.}

\section{Conclusion and outlook}
In conclusion, we study the stroboscopic properties of a time-periodic $\delta$-kicked
\textcolor{black}{$\alpha$-T$_3$ lattice. We find analylical results for its extreme limit corresponding to $\alpha=1$ i.e. dice lattice with pseudospin-1.} 
We find that with the help of the $\delta$-kicking, a variety of low energy dispersions 
including semi-Dirac type, gapless line, absolute flat quasienergy bands can be engineered around some specific points 
in the Brillouin zone. The underlying static model does not support these various types of dispersion. 
Therefore, our study can motivate the non-equilibrium transport studies on dice model given the fact that various 
quasienergy dispersion can lead to interesting transport behavior. In order to visualize the different aspects of 
quasienergy dispersion, we investigate the wave packet dynamics as a function of stroboscopic time. 
The study of wave packet dynamics reveals the existence of dynamical localization of electronic wave packet for 
a periodic kicking in the transverse direction. The dynamical localization is caused by the flat quasienergy bands. 
In particular, we obtain three absolutely flat quasienergy bands for a certain value of the kicking parameter,
namely $\lambda_z=\pi$, corresponding to the periodic kicking in the transverse direction. As a consequence, a complete
dynamical localization of wave packet is confirmed. \textcolor{black}{Additionally, the stroboscopic properties of $\alpha$-T$_3$ lattice have been studied numerically in which it is revealed that the dynamical localization of wave packet is completely absent for $0<\alpha<1$. A transverse kick also breaks the particle-hole symmetry.}

\textcolor{black}{We would like to comment that the concept of dynamical localization exists in the context of Anderson localization  in quantum kicked rotors[\onlinecite{rot2, rot4}]. The Anderson localization is a static phenomenon where all eigenstates of a system of non-interacting
particles become spatially exponentially localized in one and two dimensions in presence of disorder [\onlinecite{Ands1, Ands2}]. On the other hand, dynamical localization can be demonstrated through the coherent destruction
of tunneling for two level quantum systems[\onlinecite{TLS, TLS2}]. One can thus apparently find that dynamical localization and 
Anderson localization are two independent and disconnected physical outcomes.} 
\textcolor{black}{The quantization of conjugated variable and quantum interference restrict the energy to increase indefinitely for the periodically driven quantum systems. This 
dynamical localization in energy space eventually refers to non-ergodic behavior in phase space. This is in a way similar to the Anderson localization in real space.   }
\textcolor{black}{
Our work deals with the spatially uniform kick unlike the quantum kicked rotors with spatially dependent kicks introducing the disorder effectively  into the 
 model. Therefore, the dynamical system considered here might not be directly connected to the Anderson-like model. This being beyond the scope of the present study we leave for a future study where one can consider quenched disorder instead of the clean Hamiltonian. }

It is worthy to mention here that flat energy bands can be also obtained
when the dice lattice is subjected to a strong perpendicular magnetic field[\onlinecite{dice2}]. More specifically, 
when a $\pi$ magnetic flux per unit plaquette is piercing through the dice lattice, the resulting quasienergy bands
becomes absolutely flat. To get such flat bands one, therefore, needs a magnetic filed $B=h/(\sqrt{3}a^2 e)\sim 10^5$ T,
which is beyond the scope of present day experiments. However, an optical lattice framework provides the possibility to 
observe such effect, specifically quasienergy band engineering and dynamical localization,
experimentally using cold atoms[\onlinecite{lattice_sh}]. 
One can load Bose-Einstein condensate in the minima of an optical 
dice lattice realized by three pairs of counter propagating laser beams. By modulating the frequencies of the laser beams
the optical lattice can be made shaken in various directions which, in turn, generates a tunable 
artificial gauge field, thus emulating various strong-field physics phenomena. In fact, this artificial magnetic flux 
has been created in triangular lattices[\onlinecite{gaugeF}]. Moreover, in addition to the  optical lattice platform
we hope that our 
result can be tested in various metamaterials such as photonic [\onlinecite{Exp3,exp2,exp3}], acoustic [\onlinecite{exp4,exp5}] lattices and solid state systems [\onlinecite{Exp4}]. 

\section{Acknowledgement}
One of the authors (L. T.) sincerely acknowledges the financial supports provided by University of North Bengal through  
University Research Projects to pursue this
work.

\appendix{}
\begin{widetext}
\section{Derivation of quasienergies for a periodically kicked dice lattice}
A detailed derivation of the quasienergy spectrum of a periodically kicked dice lattice is given here.
The calculations for $x$- and $y$-kicking are similar 
and that for the $z$-kicking is less cumbersome. So, we provide only the quasienergy derivation
corresponding to $x$-kicking only.

The Floquet operator for $x$-kicking is reduced to $U_F^x(T)=e^{-i\lambda_x S_x} e^{-iH_{\bm k}T}$. We need to find out the 
eigenvalues of $U_F^x(T)$ in a straightforward manner in order to calculate the corresponding quasienergy $\Delta_{\bm k}^x$.

With the following general formula of a $3\times3$ rotation matrix
\begin{eqnarray}
U_\mu(\phi)\equiv \exp(iS_\mu\phi)=\eins-i\sin\phi\, S_\mu-(1-\cos\phi)S_\mu^2
\end{eqnarray}
with $\mu=x,y,z$,
we find
\begin{eqnarray}
\exp(-i\lambda_x S_x)=
\begin{pmatrix}\
\frac{1}{2}(1+\cos \lambda_x) & -\frac{i}{\sqrt{2}}\sin\lambda_x & \frac{1}{2}(-1+\cos \lambda_x)\\
-\frac{i}{\sqrt{2}}\sin \lambda_x & \cos \lambda_x & -\frac{i}{\sqrt{2}}\sin \lambda_x \\
\frac{1}{2}(-1+\cos \lambda_x) & -\frac{i}{\sqrt{2}}\sin \lambda_x & \frac{1}{2}(1+\cos \lambda_x)
\end{pmatrix}
=\begin{pmatrix}
 p & -iq & r\\
 -iq & 2p-1 & -iq\\
 r & -iq & p
 \end{pmatrix},
\end{eqnarray}
where $p=(1+\cos \lambda_x)/2$, $q=\sin \lambda_x/\sqrt{2}$, and $r=(-1+\cos\lambda_x)/2$.

We further find
\begin{eqnarray}
\exp(-iH_{\bm k}T)=
\begin{pmatrix}\
\frac{1}{2}[1+\cos(\omega_{\bm k} T)] &
-\frac{i}{\sqrt{2}}e^{i\theta_{\bm k}}\sin(\omega_{\bm k}T) & \frac{1}{2}e^{2i\theta_{\bm k}}[-1+\cos(\omega_{\bm k} T)]\\
-\frac{i}{\sqrt{2}}e^{-i\theta_{\bm k}}\sin(\omega_{\bm k}T) & 
\cos(\omega_{\bm k} T)] & -\frac{i}{\sqrt{2}}e^{i\theta_{\bm k}}\sin(\omega_{\bm k}T) \\
\frac{1}{2}e^{-2i\theta_{\bm k}}[-1+\cos(\omega_{\bm k} T)] & -\frac{i}{\sqrt{2}}e^{-i\theta_{\bm k}}\sin(\omega_{\bm k}T)
&\frac{1}{2}e^{-2i\theta_{\bm k}}[1+\cos(\omega_{\bm k} T)]
\end{pmatrix}
=\begin{pmatrix}
 a & -ib^\ast & c^\ast\\
 -ib & 2a-1 & -ib^\ast\\
 c & -ib & a
 \end{pmatrix},
\end{eqnarray}
where $a=[1+\cos(\omega_{\bm k}T)]/2$, $b=e^{-i\theta_{\bm k}}\sin(\omega_{\bm k}T)/\sqrt{2}$, 
and $c=e^{-2i\theta_{\bm k}}[1+\cos(\omega_{\bm k} T)]/2$. Here,
$f_{\bm k}=\omega_{\bm k} e^{i\theta_{\bm k}}$.

Now the Floquet operator can be written as 
\begin{eqnarray}
U_F^x(T)=\begin{pmatrix}
          A & -iB^\ast & C^\ast\\
          -iP & Q & -iP^\ast\\
          C & -iB & A^\ast
         \end{pmatrix},
\label{FOx}         
\end{eqnarray}
where 
\begin{center}
$A=pa-qb+rc$\\
$B=pb+q(2a-1)+rb^\ast$\\
$C=pc-qb+ra$\\
$P=(2p-1)b+q(a+c)$\\
$Q=(2p-1)(2a-1)-q(b+b^\ast)$.
\end{center}

In order to calculate the eigenvalues of $U_F^x(T)$, we find the following characteristic equation
\begin{small}
\begin{eqnarray}\label{root}
\varepsilon^3-\Big(A+A^\ast+Q\Big)\varepsilon^2+
\Bigg[\Big(A+A^\ast\Big)Q+\vert A\vert^2-\vert C \vert^2 +2{\rm Re}\Big(BP^\ast \Big)\Bigg]\varepsilon
-\Bigg[\Big(\vert A\vert^2-\vert C \vert^2\Big)Q+2{\rm Re}\Big\{B\big(AP^\ast-PC^\ast\big)\Big\}\Bigg]=0.
\end{eqnarray}
\end{small}
The roots of Eq.(\ref{root}) will give $e^{\pm i\Delta_{\bm k}T}$. 
It is straightforward to show the coefficients of $\varepsilon$ and $\varepsilon^2$ are equal and the last term will be equal to $1$.

Therefore, we have

\begin{eqnarray}
\varepsilon^3-(A+A^\ast+Q)\varepsilon^2+(A+A^\ast+Q)\varepsilon-1=0,
\end{eqnarray}
which further gives $\varepsilon=1$ and
\begin{eqnarray}
\varepsilon^2-2\kappa^x \varepsilon +1=0,
\end{eqnarray}
where $2\kappa^x=A+A^\ast+Q-1$.

Therefore, we find 
\begin{eqnarray}
\varepsilon_\pm=\kappa^x \pm \sqrt{(\kappa^x)^2-1}.
\end{eqnarray}

Note that 
\begin{eqnarray}
 \kappa^x=\frac{1}{2}\Big(\cos\lambda_x-1\Big)\sin^2\theta_{\bm k}-\sin\lambda_x
 \cos\theta_{\bm k}\sin(\omega_{\bm k}T)+
 \frac{1}{2}\Big[\sin^2\theta_{\bm k}+\cos\lambda_x\big(1+\cos^2\theta_{\bm k}\big)\Big]\cos(\omega_{\bm k}T)
\end{eqnarray}
and consequently $(\kappa^x)^2<1$.

Therefore, we have 
\begin{eqnarray}
\varepsilon_\pm=\kappa^x \pm i\sqrt{1-(\kappa^x)^2}.
\end{eqnarray}

We can now set $\varepsilon_\pm=e^{\pm i\Delta_{\bm k}^xT}$ to get the corresponding quasienergies as
\begin{eqnarray}
\Delta_{\bm k}^{x\pm}=\pm \frac{1}{T} \cos^{-1}(\kappa^x).
\end{eqnarray}

Finally, the quasienergy spectrums are obtained as
\begin{eqnarray}
\Delta_{\bm k}^{x0}=0,~~~
\Delta_{\bm k}^{x\pm}=\pm \frac{1}{T} \cos^{-1}(\kappa^x).
\end{eqnarray}

\textcolor{black}{In order to check the paticle-hole operation on the evolution operator Eq.~(\ref{FOx}), one can obtain ${\mathcal P} U_F^{x}({\bm k},T) {\mathcal P}^{-1}= U_F^{x}(-{\bm k},T)$. Here, we provide some useful information $\theta_{-\bm k}=-\theta_{\bm k}$, $a(-\bm k)=a(\bm k)=a^\ast (k)$, $b^\ast(\bm k)=b(-\bm k)$, $c^\ast(\bm k)=c(-\bm k)$, $A^\ast(\bm k)=A(-\bm k)$, $B^\ast(\bm k)=B(-\bm k)$, $C^\ast(\bm k)=C(-\bm k)$. $P^\ast(\bm k)=P(-\bm k)$ and $Q^\ast(\bm k)=Q(-\bm k)$.}

\section{Particle-Hole symmetry breaking in a kicked $\alpha$-T$_3$ lattice}
\textcolor{black}{
Here, we show explicitly how a periodic kick in a transverse direction breaks the particle-hole symmetry of a kicked $\alpha$-T$_3$ lattice for 
$0<\alpha<1$. The explicit expression of the Floquet operator corresponding to periodic kick along $z$-direction can be obtained as 
\begin{eqnarray}
 U_{F\alpha}^z(T)=\begin{pmatrix}
a_1(E\cos^2\phi+\sin^2\phi) & -ia_1F^\ast\cos\phi & a_1G^\ast\sin\phi\cos\phi \\
-ia_2F\cos\phi & a_2E & -ia_2F^\ast\sin\phi\\
a_3G\sin\phi\cos\phi& -ia_3F\sin\phi & a_3(E\sin^2\phi+\cos^2\phi)
\end{pmatrix},
\label{Faz}         
\end{eqnarray}
where $a_1=\exp(-2i\lambda_z\cos^2\phi)$, 
$a_2=\exp(2i\lambda_z\cos(2\phi))$, 
$a_3=\exp(2i\lambda_z\sin^2\phi)$, 
$E=\cos(\omega_{\bm k}T)$, $F=e^{-i\theta_{\bm k}}\sin(\omega_{\bm k}T)$, and $G=e^{-2i\theta_{\bm k}}[\cos(\omega_{\bm k}T)-1]$.
It is readily evident from Eq.(\ref{Faz}) that 
${\mathcal P} U_{F\alpha}^{z}({\bm k},T) {\mathcal P}^{-1}\neq U_{F\alpha}^{z}(-{\bm k},T)$ provided $a_{1,2,3}$ are imaginary. This is reflected in specific form of $\beta $ Eq.~(\ref{beta_general}) which becomes imaginary for $z$-kick.}

\end{widetext}

\end{document}